\documentstyle{mn}
\input{epsf}
\newif\ifAMStwofonts

\def\lesssim{\mathrel{\hbox{\rlap{\hbox{\lower4pt\hbox{$\sim$}}}\hbox{$<$}}}}

\def\gtrsim{\mathrel{\hbox{\rlap{\hbox{\lower4pt\hbox{$\sim$}}}\hbox{$>$}}}}

\def\msun{M$_{\odot}$}

\def\lteff{$\log{T_{\rm eff}}$~}

\def\ll_lsun{Log$({L/\rm L_{\odot}})$~}

\def\masa_msun{$M/ \rm M_{\odot}$~}

\def\m_mstar{$M/M_{*}$~}

\title[Evolution of a 3 \msun \ star from the main sequence to the
ZZ Ceti stage]{Evolution of a 3 \msun \ star from the main sequence to the
ZZ Ceti stage: the role played by element diffusion}

\author[L. G. Althaus, A. M. Serenelli, A. H. C\'orsico and O. G. Benvenuto]
{L.   G. Althaus\thanks{Member of the Carrera del Investigador
Cient\'{\i}fico y Tecnol\'ogico, Consejo Nacional de
Investigaciones Cient\'{\i}ficas y T\'ecnicas (CONICET), Argentina.},
A. M. Serenelli\thanks{Fellow of CONICET, Argentina.},
A. H. C\'orsico\thanks{Fellow of CONICET, Argentina.} and
O. G. Benvenuto\thanks{Member  of  the  Carrera  del Investigador
Cient\'{\i}fico, Comisi\'on  de Investigaciones  Cient\'{\i}ficas
de la Provincia de Buenos Aires, Argentina.}\\
Facultad  de  Ciencias
Astron\'omicas y Geof\'{\i}sicas, Universidad Nacional de La
Plata, Paseo del Bosque S/N, (1900) La Plata, Argentina\\
E-mail:althaus,serenell,acorsico,obenvenu@fcaglp.fcaglp.unlp.edu.ar}

\pagerange{\pageref{firstpage}--\pageref{lastpage}}

\pubyear{2001}

\begin{document}

\maketitle

\label{firstpage}

\begin{abstract}

The  purpose  of  this  paper  is to  present  new  full  evolutionary
calculations for DA white dwarf  stars with the major aim of providing
a  physically  sound  reference  frame  for  exploring  the  pulsation
properties of  the resulting  models in future  communications.  Here,
white dwarf  evolution is followed  in a self-consistent way  with the
predictions of  time dependent element diffusion  and nuclear burning.
In addition, full account is taken of the evolutionary stages prior to
the white dwarf formation.  In  particular, we follow the evolution of
a  3  \msun  model  from  the  zero-age  main  sequence  (the  adopted
metallicity is  Z=0.02) all  the way from  the stages of  hydrogen and
helium burning in  the core up to the  thermally pulsing phase.  After
experiencing 11 thermal pulses, the  model is forced to evolve towards
its   white  dwarf   configuration  by   invoking  strong   mass  loss
episodes. Further evolution  is followed down to the  domain of the ZZ
Ceti stars on the white dwarf cooling branch.

Emphasis  is  placed  on  the  evolution  of  the  chemical  abundance
distribution  due  to  diffusion  processes  and the  role  played  by
hydrogen  burning during  the  white dwarf  evolution.   We find  that
discontinuities  in the  abundance distribution  at the  start  of the
cooling branch are considerably smoothed out by diffusion processes by
the time  the ZZ  Ceti domain is  reached. Nuclear burning  during the
white  dwarf stage does  not represent  a major  source of  energy, as
expected for a progenitor star of initially high metallicity.  We also
find  that thermal diffusion  lessens even  further the  importance of
nuclear burning.

Furthermore, the implications of  our evolutionary models for the main
quantities   relevant    for   adiabatic   pulsation    analysis   are
discussed.  Interestingly, the shape  of the  Ledoux term  is markedly
smoother as  compared with previous detailed studies  of white dwarfs.
This   is    translated   into   a   different    behaviour   of   the
Brunt-V\"ais\"al\"a frequency.

\end{abstract}

\begin{keywords}  stars:  evolution  -  stars: interiors - stars:
white dwarfs - stars: oscillations

\end{keywords}

\section{INTRODUCTION} \label{sec:intro}

Over the  last few years,  radial and particularly  non-radial stellar
pulsations have  become a  very powerful tool  for inquiring  into the
internal structure  and evolution of stars.  Thanks  to the increasing
degree  of  sophistication   both  in  theoretical  and  observational
techniques, asteroseismology has been successfully applied to decipher
the oscillatory pattern of  numerous pulsating stars, amongst them our
Sun represents the best  example.  With the advancement and refinement
of  observations, a  large  number  of stellar  objects  located in  a
variety  of places  in the  Hertzsprung-Russell diagram  has gradually
manifested themselves in non-radial pulsators.  Indeed, variable stars
covering  several evolutionary  stages,  such as  roAp, SPB,  $\delta$
Scuti, $\beta$  Cephei and variable white dwarfs  have been classified
as non-radial pulsators (see, e.g.,  Cox 1980; Unno et al. 1989; Brown
\& Gilliland  1994 and  Gaustchy \& Saio  1995, 1996).  To  this class
belong also  the recently discovered  sdB (Kilkenny et al.   1997) and
$\gamma$ Doradus (Kaye et al. 1999) variable stars.

From the  observational point of  view (with the obvious  exception of
our  Sun), white  dwarfs represent  one  of the  best established  and
extensively  studied  kind of  non-radial  pulsators. Pulsating  white
dwarfs   exhibits  multi-periodic   luminosity  variations   in  three
different regions of  the Hertzsprung-Russell diagram corresponding to
the  currently called  DOV (and  PNNV), DBV  and DAV  (see,  e.g., the
review by Winget  1988).  Of particular interest in  this work are the
DAVs (hydrogen-rich atmospheres), or ZZ  Ceti stars, that are found to
pulsate  in  the  instability  strip corresponding  to  the  effective
temperature  ($T_{\rm eff}$)  range of  12500 K  $\geq$  $T_{\rm eff}$
$\geq$ 10700 K.  The periodicities in their light curves are basically
explained in terms of non-radial g-modes of low harmonic degree ($\ell
\leq 2$), driven by the $\kappa-\gamma$ mechanism working in a partial
ionization region  below the stellar surface (Dolez  \& Vauclair 1981;
Winget et al.  1982)  \footnote{However, Brickhill (1991) proposed the
convective   driving   mechanism   as   being  responsible   for   the
overstability of  g-modes in DAVs  (see also Goldreich \&  Wu 1999).}.
The  periods are  found within  a range  of 100-1200s  and photometric
amplitudes reach  up to 0.30  magnitudes.  Numerous studies  have been
devoted  to  analysing  the  pulsation characteristics  of  DAV  white
dwarfs.  Amongst  them, we mention  the works by Tassoul,  Fontaine \&
Winget  (1990), Brassard  et al.   (1991, 1992ab),  Bradley  \& Winget
(1994), Gautschy,  Ludwig \& Freytag  (1996) and Bradley  (1996, 1998,
2001).

In order to  fully understand the oscillatory properties  of DAV white
dwarfs and to take full advantage of the richness offered by available
observations, physically  sound stellar  models are required.   In the
context of these pulsating stars,  most of the existing research rests
on  stellar   models  constructed   under  the  assumptions   of  some
simplifying hypothesis.  This is  particularly true with regard to the
treatment  of the  chemical  abundance distribution.   In this  sense,
carbon and oxygen profiles are  usually treated as free parameters. In
addition, the  chemical abundance distribution is assumed  to be fixed
during the evolution  across the instability domain and  in some cases
the diffusive equilibrium approximation is invoked to assess the shape
of the chemical profile at the interface regions.  Hydrogen burning is
likewise neglected in the construction of evolutionary models employed
in  pulsation studies  of  ZZ  Ceti stars.   The  neglect of  hydrogen
burning is only justified if the white dwarf is formed with relatively
thin hydrogen  layers.  However, if  the hydrogen envelope  is massive
enough, hydrogen burning reactions constitute an appreciable source of
energy  even during the  evolutionary stages  corresponding to  the ZZ
Ceti domain (Iben  \& Tutukov 1984).  The details  of burning are more
complex  if element diffusion  is allowed  to operate.   Indeed, white
dwarf  evolutionary  calculations  in  which  time  dependent  element
diffusion is properly accounted for (Iben \& MacDonald 1986) show that
nuclear burning via the CN  reactions plays a different role depending
on  the mass  of the  helium buffer  region between  the hydrogen-rich
envelope  and  the carbon-  and  helium-rich  underlying layers.   For
instance,  a diffusion-induced  hydrogen  shell flash  is expected  to
occur if the  helium buffer is sufficiently less  massive. How massive
the helium buffer can be depends critically on the phase in the helium
shell  flash cycle  during the  thermally pulsing  stage at  which the
progenitor  departs  from  the  asymptotic  giant  branch  (AGB)  (see
D'Antona \& Mazzitelli 1990 for details).

Obviously,  the construction  of stellar  models of  DAV  white dwarfs
appropriate for pulsation studies  in which the above mentioned issues
are  fully  taken  into  account  requires  evolutionary  calculations
considering  not only  time  dependent element  diffusion  but also  a
detailed treatment of the evolutionary stages prior to the white dwarf
formation.  The calculation of such stellar models is the main purpose
of the present work and to the best of our knowledge such an endeavour
has never  been attempted.   The primary application  of the  DA white
dwarf evolutionary models to be presented here will be the exploration
of their pulsation properties in future papers.  Specifically, in this
paper we shall limit ourselves to discuss the evolutionary results and
their  implications for  the  main quantities  entering the  adiabatic
pulsation equations.

White  dwarf evolution  treated  in a  self-consistent  way with  time
dependent  element diffusion  is an  important aspect  of  the present
work. In most of previous pulsation studies, the equilibrium diffusion
in the trace element approximation  has been used to specify the shape
of  the  chemical profile  at  the  composition  transition region  of
evolving  stellar  models  (see  Tassoul  et al.   1990;  Brassard  et
al. 1991,  1992ab; Bradley 1996,  1998, 2001; Bradley \&  Winget 1994;
Montgomery,  Metcalfe  \&  Winget  2001).   However,  the  equilibrium
approach for  diffusion is  not valid when  diffusion time  scales are
comparable  to  the   evolutionary  ones.   In  particular,  diffusive
equilibrium in  the deep  layers of  the white dwarf  model is  not an
adequate approximation even at the  ages characteristic of the ZZ Ceti
stage.   In fact,  it  is  found that  during  such stages,  diffusion
modifies the spatial distribution of the elements, particularly at the
chemical  interfaces  (see Iben  \&  MacDonald  1985).   For a  proper
treatment of the diffusively  evolving stratifications, we consider in
this  work  the  processes  of gravitational  settling,  chemical  and
thermal  diffusion  following  the  treatment of  Burgers  (1969)  for
multicomponent gases.  In the context of DA white dwarf evolution, the
treatment for diffusion we shall use here has been employed by Iben \&
MacDonald (1985,  1986).  In addition,  Dehner \& Kawaler  (1995) have
used non-equilibrium diffusion star models for studying the connection
between  DO  and  DB  white  dwarfs.  The  shape  of  the  composition
transition  zone  is a  matter  of the  utmost  importance  as far  as
asteroseismology is  concerned.  In particular, it  contributes to the
shape  of  the  Ledoux   term  appearing  in  the  Brunt-V\"ais\"al\"a
frequency (Brassard  et al.   1991) and plays  a critical role  in the
phenomenon of mode trapping in white dwarfs (see Tassoul et al.  1990;
Brassard et al.  1992a and references cited therein).

Another important aspect of the present study is that the evolutionary
stages  prior  to the  white  dwarf  formation  are fully  taken  into
account.   Specifically, we started  our calculations  from a  3 \msun
stellar model at  the zero-age main sequence (ZAMS)  and we follow its
further evolution  all the way from  the stage of  hydrogen and helium
burning in  the core  up to the  tip of  the AGB where  helium thermal
pulses  occur. To assure  a full  relaxation of  the helium  zone (see
Mazzitelli  \& D'Antona  1986),  we computed  a  reasonable number  of
thermal pulses, after which the progenitor is forced to evolve towards
its white  dwarf configuration by invoking strong  mass loss episodes.
We shall concentrate on the  particular situation that the white dwarf
progenitor  departs  from  the  AGB  when  stationary  helium  burning
primarily supports  the star  luminosity, following the  occurrence of
the last  helium thermal  pulse. We have  not invoked  additional mass
losses during the  planetary nebula stage or early  during the cooling
branch. This will  allow us to examine the maximum  mass value for the
remaining  hydrogen content  as predicted  by the  particular  case of
evolution  analysed in  this work,  enabling  us to  explore the  role
played by nuclear burning during the cooling stages.

The  paper is  organised as  follows.  In  Section 2  we  describe our
evolutionary   code  and   the   treatment  we   follow  for   element
diffusion. In Section 3 we  present in detail the evolutionary results
for  the white  dwarf progenitor,  giving particular  emphasis  to the
thermal micro-pulses appearing towards  the end of core helium burning
and the  thermally pulsing phase  during the final AGB  evolution.  In
this section, we  also present the results for  the post-AGB and white
dwarf evolution.  Attention is focused  mainly on the evolution of the
chemical distribution resulting from  diffusion processes and the role
of nuclear burning during the white dwarf regime.  The implications of
the  evolutionary  results  for   the  main  quantities  relevant  for
adiabatic  pulsation analysis  are  detailed in  Section 4.   Finally,
Section 5 is devoted to summarizing our results.

\section{COMPUTATIONAL DETAILS}

The calculations presented  in this work have been  performed with the
stellar evolutionary code developed  at La Plata Observatory. The code
has  been  used in  our  previous  studies  on white  dwarf  evolution
(Althaus \& Benvenuto  1997, 2000; Benvenuto \& Althaus  1998), and it
has  recently been appropriately  modified in  order to  calculate the
evolutionary stages prior to the  formation of white dwarfs.  In broad
outline, the  code is  based on the  method of Kippenhahn,  Weigert \&
Hofmeister (1967)  for calculating stellar  evolution.  In particular,
to specify the outer boundary conditions we carried out three envelope
integrations  from photospheric  starting values  inward to  a fitting
outer mass  fraction, which is  located in mass near  the photosphere.
The interior  solution is obtained via the  canonical Henyey iteration
scheme  as described  by Kippenhahn  et al.   (1967).  To  improve the
numerical stability of  our code, the Henyey scheme  is applied to the
differences in  the physical quantities  (luminosity, pressure, radius
and  temperature) between the  previous and  the computed  models.  In
addition, models were divided  into approximately 1500 mesh points and
time  steps were maintained  small enough  so as  to get  a reasonable
numerical accuracy  during the thermal pulses  at the tip  of the AGB.
Mesh  distribution is  regularly updated  every five  time  steps. Our
algorithm  inserts mesh  points where  they are  most needed,  that is
where physical variables change appreciably, and eliminates them where
they  are  not  neccesary.   We   want  to  mention  that  the  entire
evolutionary sequence from the ZAMS to the white dwarf stage comprises
about 60000 stellar models.

The   constitutive   physics   are   as  detailed   and   updated   as
possible.  Briefly, it comprises  OPAL radiative  opacities (including
carbon-  and   oxygen-rich  opacities)  for   arbitrary  metallicities
(Iglesias \&  Rogers 1996), complemented at low  temperatures with the
molecular opacities from Alexander  \& Ferguson (1994). In particular,
opacities  for varying  metallicities  are required  during the  white
dwarf  regime when  account is  taken in  the calculations  of element
diffusion.   Conductive opacity  for the  high-density regime  is from
Itoh et  al.  (1983)  and Hubbard \&  Lampe (1969) for  low densities.
Neutrino  emission rates  for pair,  photo, plasma  and Bremsstrahlung
processes have been taken into account according to the formulation of
Itoh and  collaborators (see Althaus  \& Benvenuto 1997  for details).
As far as the equation of state is concerned, we have included partial
ionization,   radiation  pressure,   ionic   contributions,  partially
degenerate electrons  and Coulomb  interactions.  For the  white dwarf
regime, we employ an updated version of the equation of state of Magni
\&   Mazzitelli  (1979).   We   have  considered   a  network   of  30
thermonuclear  reaction rates for  hydrogen burning  (corresponding to
the  proton-proton chain  and the  CNO bi-cycle)  and  helium burning.
Nuclear reaction rates are taken from Caughlan \& Fowler (1988) except
for the  reaction $^{12}{\rm C}(\alpha, \gamma)^{16}{\rm  O}$ which is
taken from Angulo et al. (1999)  (this rate is about twice as large as
that of Caughlan \& Fowler 1988).  Electron screening is from Graboske
et al.  (1973)  and Wallace, Woosley \& Weaver  (1982).  The change in
the chemical composition resulting from nuclear burning is computed by
means of a standard implicit  method of integration. In particular, we
follow  the  evolution  of  the chemical  species  $^{1}$H,  $^{3}$He,
$^{4}$He, $^{7}$Li, $^{7}$Be,  $^{12}$C, $^{13}$C, $^{14}$N, $^{15}$N,
$^{16}$O,  $^{17}$O,  $^{18}$O  and  $^{19}$F.   Convection  has  been
treated  following the standard  mixing length  theory (B\"ohm-Vitense
1958) with mixing-length to pressure  scale height $\alpha= $1.5.  The
Schwarzschild  criterium  was  used  to determine  the  boundaries  of
convective  regions.    Overshooting  and  semi-convection   were  not
considered.

In this  work, we follow  the evolution of  an initially 3  \msun star
starting at  the ZAMS.  The adopted  metallicity $Z$ is  $Z$= 0.02 and
the   initial  abundance  by   mass  of   hydrogen  and   helium  are,
respectively,   $X_{\rm   H}$=   0.705   and   $X_{\rm   He}$=   0.275
(representative  for  solar values  as  given  by  Anders \&  Grevesse
1989). Evolution  has been computed  at constant stellar mass  all the
way from the  stages of hydrogen and helium burning in  the core up to
the tip  of the AGB where  helium thermal pulses occur.   To achieve a
white  dwarf   configuration,  an   artificial  mass  loss   has  been
incorporated in  our evolutionary  code.  Specifically, mass  loss was
initiated when the white dwarf  progenitor was about to experience its
eleventh  thermal pulse.   The adopted  mass loss  rate  was $10^{-4}$
\msun yr$^{-1}$ and it was  applied to each stellar model as evolution
proceeded.  After the convergence of each new stellar model, the total
stellar mass is reduced according to  the time step used for the model
and the  mesh points are  appropriately adjusted.  We want  to mention
that  because  of  the  high  computational demands  involved  in  the
calculation we  perform here, particularly regarding  the treatment of
white  dwarf  evolution  with  diffusively  evolving  abundances  (see
below), we shall restrict ourselves exclusively to examine one case of
evolution for  the white dwarf  progenitor, deferring a  more detailed
exploration  of different  possibilities of  white dwarf  formation to
future works.

The evolution of the  chemical abundance distribution due to diffusion
processes during  the whole white dwarf stage  represents an important
aspect of the present study.  In our treatment of element diffusion we
have  considered  gravitational  settling,  and chemical  and  thermal
diffusion of nuclear  species.  To this end, we  adopted the treatment
for multicomponent gases presented by Burgers (1969), avoiding the use
of  the trace  element approximation  usually assumed  in  white dwarf
studies. It is worth noting that when the progenitor star departs from
the AGB after  the end of mass loss episodes, its  envelope is made up
of a mixture of hydrogen and helium.  Also, in deeper layers and below
the nearly pure  helium buffer, there is an  underlying region rich in
both helium and  carbon (see next section); thus the  use of the trace
element approximation  would not be  appropriate for our  purposes. In
this study we are interested in the chemical evolution occurring quite
deep in  the star,  thus radiative levitation  and possible  wind mass
loss from  the surface  during the hot  white dwarf stages,  which are
expected to  alter the surface  composition of these stars,  have been
neglected (see Unglaub \& Bues 2000 for a recent detailed study of the
evolution  of  chemical abundances  in  surface  layers  of hot  white
dwarfs).  In the  context of white dwarf evolution,  the treatment for
diffusion we  use here has been  employed by Iben  \& MacDonald (1985,
1986) (thermal diffusion  not included).  Recently, it  was applied by
MacDonald, Hernanz \&  Jos\'e (1998) to address the  problem of carbon
dredge-up in  white dwarfs with helium-rich envelopes  and by Althaus,
Serenelli  \&  Benvenuto  (2001a,b)  to  explore the  role  played  by
diffusion in  inducing thermonuclear flashes  in low-mass, helium-core
white dwarfs.

Details about the procedure we follow to solve the diffusion equations
are  in Althaus  \& Benvenuto  (2000).  In  particular, we  follow the
evolution  of  the  isotopes  $^{1}$H, $^{3}$He,  $^{4}$He,  $^{12}$C,
$^{14}$N and  $^{16}$O. In  order to calculate  the dependence  of the
structure  of  our white  dwarf  models  on  the diffusively  evolving
abundances   self-consistently,  the   set  of   equations  describing
diffusion has been coupled  to our evolutionary code.  After computing
the  change of  abundances by  effect of  diffusion, they  are evolved
according  to the  requirements  of nuclear  reactions and  convective
mixing.   Finally, we  emphasize that  radiative opacities  during the
white dwarf  regime are  calculated for metallicities  consistent with
the diffusion predictions.  In particular, the metallicity is taken as
two  times the  abundances of  CNO elements  as suggested  by  Iben \&
MacDonald (1986).

\section{EVOLUTIONARY RESULTS} \label{sec:results}

\subsection{White dwarf progenitor}

Here  we  describe  the  results   we  obtained  with  regard  to  the
evolutionary phases prior to the white dwarf formation. We shall limit
ourselves to describe the main features of such evolution particularly
those which are of immediate  relevance for the white dwarf formation,
and we refer the reader to the works of Mazzitelli \& D'Antona (1986),
Vassiliadis  \& Wood (1993),  Bl\"ocker (1995a)  amongst others  for a
more complete description about  the evolution of low and intermediate
mass stars.  We begin by  examining the complete evolutionary track in
the  Hertzsprung-Russell diagram  that is  illustrated in  Fig.1.  Our
numerical  simulation  covers  all   the  evolutionary  phases  of  an
initially 3 \msun star from the ZAMS to the domain of ZZ Ceti stars on
the white dwarf  cooling branch. The age (in units  of $10^4$ yr) from
the end  of mass loss episodes and  the mass of hydrogen  (in units of
$10^{-4}$  \msun) are indicated  at selected  points along  the track.
For clarity,  the evolutionary phases  corresponding to mass  loss are
not plotted.

After $4.1  \times 10^{8}$ yr  of evolution and  by the end  of helium
burning in the core, the  first feature worthy of comment predicted by
our calculations is the appearance of a series of micro-pulses (not to
be confused with the major thermal pulses on the AGB) of low amplitude
in  the  surface luminosity.  Such  micro-pulses  are  due to  thermal
instabilities  in  the  helium-burning  shell above  the  carbon-  and
oxygen-  rich core,  which  causes the  helium  luminosity to  undergo
oscillations.   Specifically,  such  pulses  appear when  the  central
helium  abundance  by  mass  falls  below  $\approx$  0.001  (and  the
convective  core  vanishes).   The  time  dependence  of  the  surface
luminosity $L_*$  and the  helium-burning luminosity $L_{\rm  He}$ (in
solar units)  during the micro-pulse phase  is shown in  two insets in
Fig.  1, where the time scale is given in million years from the ZAMS.
A total of 40 micro-pulses  with an inter-pulse period of $\approx 3.1
\times  10^{5}$ yr  occurred.  Note  that the  pulse amplitude  in the
surface luminosity is indeed  modest ($\Delta$ \ll_lsun $\approx$ 0.02
at most).  It is  worth mentioning that thermal micro-pulses occurring
in a  3 \msun star  towards the end  of its core  helium-burning phase
have also been  reported by Mazzitelli \& D'Antona  (1986).  After the
end of micro-pulses, the star evolves until complete helium exhaustion
at  the centre.  The total  time spent  during central  helium burning
amounts to about $1.25 \times 10^{8}$ yr.

After  helium is  exhausted  in  the core,  leaving  a central  oxygen
abundance of 0.62 by mass, evolution proceeds towards the phase of the
major thermal pulses  on the AGB, at which  point helium shell burning
becomes unstable  again. The AGB evolution  of intermediate-mass stars
is well known to be characterized by helium shell flashes during which
the  burning rate  rises very  steeply.  In  our simulation,  the time
elapsed from  central helium exhaustion until the  first thermal pulse
is  $2.6  \times  10^{7}$  yr,  when the  surface  luminosity  exceeds
\ll_lsun= 3.2 for the first time in its evolution.  After experiencing
11 thermal pulses and considerable mass loss, the mass of the hydrogen
envelope is  reduced so much  that the star  departs from the  AGB and
evolves towards  large effective temperatures.  This  takes place when
the  star luminosity is  supported by  stationary helium  burning.  We
will discuss the implications of this situation later in this section,
for  now suffice  it to  say  that when  mass loss  ends, the  remnant
consumes  a  considerable fraction  of  its  remaining  hydrogen in  a
redward  ``hook''  on the  Hertzsprung-Russell  diagram.  During  this
phase,  evolution proceeds  very slowly  \footnote{We want  to mention
that  we have  not considered  further mass  loss after  the  star has
reached \lteff= 3.8  for the first time after leaving  the AGB.}. As a
result of mass loss episodes, the stellar mass has been decreased from
3 to  0.563~ \msun.  Eventually,  the remnant reaches the  white dwarf
cooling  branch.  Thereafter,  element diffusion  alters  the chemical
abundance  distribution within the  star even  at the  lowest computed
luminosity stages.  During the  white dwarf cooling phase, the coupled
effects of hydrogen  burning and element diffusion reduce  the mass of
hydrogen that is left in outer layers by almost a factor of 2.

The  time dependence of  the surface  luminosity during  the thermally
pulsing phase is detailed in the upper panel of Fig. 2, where the time
scale is given in million years  counted from the ZAMS.  A total of 11
thermal pulses  have been computed  before the white  dwarf progenitor
departs from  the AGB.   This phase of  evolution has been  studied in
detail by numerous authors in the literature (Sch\"onberner 1979; Iben
1982; Iben \&  Renzini 1983; Vassiliadis \& Wood  1993 amongst others)
and we refer the reader to  those studies for details.  Let us however
mention  some  words  about   the  role  played  by  various  relevant
luminosities by the time the star  is next to evolve away from the AGB
towards the  white dwarf  state.  To  this end, we  show in  the lower
panel of Fig.   2 the evolution of the  surface luminosity ($L_*$) and
the hydrogen-and helium-burning  luminosities ($L_{\rm H}$ and $L_{\rm
He}$ respectively)  during and between the tenth  and eleventh pulses.
Departure from  the AGB could  in principle occur at  different stages
during the inter-pulse.

Here, we will  direct our attention to one  particular situation: that
in which the  star leaves the AGB during  the quiescent helium-burning
phase following the eleventh  helium thermal pulse.  Our simulation is
thus  representative of the  possibility that  departure from  the AGB
takes place  early in the  helium shell flash cycle.   The consequence
for  the  post-AGB evolution  resulting  from  departure at  different
locations on  the inter-pulse has been carefully  explored by numerous
investigators,  amongst them  Iben  (1984), Wood  \& Faulkner  (1986),
Bl\"ocker (1995b). Such studies  show for instance that the transition
time from AGB  to the planetary nebula region  depends strongly on the
phase at  which the  star leaves  the AGB.  From  the bottom  panel of
Fig. 2  we see that  the star departs  from the AGB when  helium shell
burning   is  dominant.    Specifically,   the  helium-burning   shell
contributes 82~\% of the surface  luminosity by the time the effective
temperature begins to increase. In  the meantime the hydrogen shell is
almost extinguished.  It is  worthwhile to comment that helium-burning
luminosity in the shell source reaches $5 \times 10 ^6 \rm L_{\odot}$,
causing an  expansion in  the layers above  with the  consequence that
hydrogen burning decreases significantly to be almost extinguished and
leading  to a  sharp spike  in  the surface  luminosity.  After  that,
helium luminosity begins to drop and hydrogen is re-ignited.

For the sake  of completeness, we show in Fig. 3  the behaviour of the
central  conditions  during  the  whole evolution  from  the  hydrogen
burning in  the core on  the main sequence  to the white  dwarf state.
Relevant episodes in the life of the star are indicated in the figure.
In particular, the response of the central region to the occurrence of
both  thermal pulses  and  micro-pulses is  noted.   Once the  remnant
leaves the AGB, evolution  proceeds at almost constant central density
to become a white dwarf.

\subsection{Post-AGB and white dwarf evolution}

After the end of the thermally pulsing phase, the remnant star departs
from  the asymptotic  giant branch  at low  effective  temperature and
evolves  as  a  planetary   nebula  nuclei  towards  higher  effective
temperatures.   In  Fig.   4  we  show  as  a  function  of  effective
temperature  the  mass  of  hydrogen  in outer  layers  (in  units  of
$10^{-3}$  \msun) for the  0.563 $M_{\odot}$  post-AGB remnant  from a
stage in  the evolution  just before  the end of  mass loss  till that
corresponding  to  the ZZ  Ceti  domain  in  the white  dwarf  regime.
Numbers in parenthesis besides circles at selected points on the curve
give  the age  in years  counted from  the end  of mass  loss  and the
contribution of  the helium shell  burning to surface  luminosity.  By
the end of  mass loss (at \lteff= 3.8) the  contribution of the helium
shell  burning  to surface  luminosity  has  been  reduced to  55  \%.
Thereafter,  the remnant  star returns  to the  AGB and  increases its
surface luminosity  up to the inter-flash maximum  value.  During this
phase, helium burning becomes  less important and the hydrogen content
in  outer  layers is  reduced,  as  a  result of  increasing  hydrogen
burning, from $\approx 2 \times 10^{-3}$ \msun at the end of mass loss
down to $\approx 5 \times  10^{-4}$ \msun once the remnant resumes its
evolution to  the blue. By the  time the mass of  the hydrogen content
decreases  below  $\approx 8  \times  10^{-4}$  \msun, helium  burning
becomes virtually extinct and  hydrogen burning via CN cycle reactions
becomes the dominant source of surface luminosity.  It is worth noting
that evolution proceeds very slowly  during this phase and the remnant
star  spends a substantial  fraction of  its post-AGB  transition time
there (about 46000 yr).  Note also  that the star takes about 48000 yr
to  reach  an  effective   temperature  of  30000K  required  for  the
excitation  of the planetary  nebula.  This  time is  so long  that no
planetary  nebula is produced.   These results  qualitatively resemble
those of  Mazzitelli \& D'Antona  (1986), who found that  if departure
from the  AGB takes place  during the quiescent helium  burning phase,
then, when  mass loss  is stopped, the  remnant spends  a considerable
time in consuming  most of its hydrogen envelope  as a red supergiant.
Long  post-AGB evolutionary  time scales  were also  found by  Wood \&
Faulkner (1986) when their models  are forced to abandon the AGB early
in the inter-pulse phase.  As  well known, post-AGB times are strongly
dependent on  the phase of the  helium shell flash cycle  at which the
star leaves the  AGB.  In this regard, had  our white dwarf progenitor
been forced  to depart from the  AGB somewhat later  than assumed here
while burning  hydrogen (by  employing a smaller  mass loss  rate), it
would  not have  returned  to the  AGB\footnote{Should departure  have
occurred very late in the pulse  cycle, then a last helium shell flash
would be  expected to  occur in the  planetary nebula nuclei  stage or
even in the  white dwarf regime, thus giving rise  to a born-again AGB
star (Sch\"onberner 1979; Iben 1984).} and its further evolution would
thus have proceeded  much faster (see Mazzitelli \&  D'Antona 1986 and
Wood \& Faulkner  1986).  Note finally that the  mass of hydrogen that
is left in outer layers at the  start of the cooling branch is about $
1.5 \times 10^{-4}$  \msun, and this is reduced  to $7 \times 10^{-5}$
\msun by the time the ZZ  Ceti domain is reached.  Because we have not
invoked  additional mass  loss  episodes during  the planetary  nebula
stage or  early during  the cooling branch,  the quoted value  for the
final hydrogen mass should be considered as an upper limit.  We stress
that in the present calculation we have adopted an initial metallicity
of $Z=  0.02$; much lower  values of $Z$  would give rise to  a larger
final  hydrogen envelope  mass  for  the same  stellar  mass (see  for
instance Iben \& MacDonald 1986).

Since the chemical  stratification of the white dwarf  is relevant for
our  purposes, let  us detail  the resulting  chemical profile  of the
post-AGB  remnant.  In  particular,  the carbon  and oxygen  abundance
distribution within the core of the remnant is detailed in Fig. 5. The
inner  part  of  the  core  of  carbon and  oxygen  emerges  from  the
convective helium core burning and from the subsequent stages in which
the helium-burning  shell propagates outwards. Note that  this kind of
profile is  the typical one,  apart from differences arising  from the
employment  of different cross  sections for  the $^{12}$C  $+ \alpha$
reaction   rate,   predicted   by   evolutionary   models   in   which
semi-convection and  overshooting are  not considered.  The  degree to
which  semi-convection  and  overshooting  affect  the  core  chemical
stratification is an issue subject to debate and in most of studies in
which this aspect is addressed, they are treated very roughly. Here we
prefer not to  include any of them in our models  and this fact should
be kept in mind by the  reader.  We simply want to mention that larger
oxygen abundances are expected if convective overshooting brings fresh
helium into the  core during the final helium  burning phases (see for
instance Mazzitelli  \& D'Antona 1986  for more details).   Before the
remnant reaches its white dwarf configuration, a mixing episode occurs
towards the  central regions of the  star. Indeed, we find  that, as a
result  of  the  particular  shape  of the  carbon-oxygen  profile  at
$M_r/M_*  \approx 0.2$, a  Rayleigh-Taylor instability  develops which
gives rise to a chemical rehomogenization of the innermost zone of the
star (see  Salaris et al. 1997  for a similar  finding). The resulting
carbon-oxygen distribution  after rehomogenization, which  is shown in
Fig.  5  with thick  lines, is the  adopted one  in this work  for the
remaining  white  dwarf   evolution.   Surrounding  the  carbon-oxygen
interior there  is a  shell rich in  both carbon ($\approx$  35\%) and
helium ($\approx$  60\%), and an overlying layer  consisting of nearly
pure helium,  the so-called helium buffer  (see the inset  in Fig. 5).
The  presence of  carbon in  the helium-rich  region below  the helium
buffer stems  from the short-lived convective mixing  which has driven
the carbon-rich zone upwards during  the peak of the last helium pulse
on  the  AGB.  The  mass  of  the helium  buffer  is  of relevance  in
connection with the further evolution  of the remnant during the white
dwarf state, particularly regarding the occurrence of a hydrogen shell
flash induced by element diffusion.  In this regard, Iben \& MacDonald
(1986) have found that if the helium  buffer mass is as small as 0.001
\msun,  then  a hydrogen  shell  flash is  initiated  as  a result  of
chemical diffusion.   In our  simulation, when the  progenitor departs
from the  AGB, the mass of  the helium buffer is  $3.4 \times 10^{-4}$
\msun, but during the evolutionary phases in which the star, after the
end of  mass loss  episodes, returns to  the AGB  to burn most  of the
remaining  hydrogen content, the  buffer mass  increases up  to 0.0024
\msun (and  to 0.003 \msun  at the white  dwarf birth) as a  result of
nuclear burning at the base of the hydrogen envelope. The final helium
buffer has thus become massive enough that a hydrogen shell flash will
not occur  at advanced  stages of evolution  (see below).  We  want to
stress that  while the buffer  mass is extremely small  when departure
from AGB takes  place during quiescent helium burning,  the burning of
all excess hydrogen in the red prevents the existence of helium buffer
masses as low  as $\approx$ 0.001 \msun from  occurring. This inhibits
the occurrence of diffusion-induced  hydrogen shell flashes during the
white dwarf phase. We speculate that small helium buffer masses at the
start of the  cooling branch could be possible only  in cases when the
progenitor leaves the AGB during the helium shell flash at the surface
luminosity peak or during the early portion of the stationary hydrogen
burning  phase.  We   plan  to  place  this  speculation   on  a  more
quantitative basis in a future work.  Finally, we want to mention that
the total helium content within  the star once helium shell burning is
eventually extinguished remains 0.014 \msun.

The  time dependence of  the luminosity  contribution due  to hydrogen
burning via  proton-proton reactions  ($L_{\rm pp}$) and  CNO bi-cycle
($L_{\rm CNO}$),  helium burning  ($L_{\rm He}$), and  neutrino losses
($L_{\nu}$), as well  as the surface luminosity ($L_*$)  for the 0.563
$M_{\odot}$ white dwarf remnant is shown  in the upper panel of Fig. 6
from evolutionary stages corresponding  to the planetary nebula nuclei
down to the lowest  computed surface luminosities. Luminosities are in
solar units, and the age is counted in years from the moment the model
reaches  the  point  defined  by  $\log{L/\rm  L_{\odot}}$=  3.71  and
$\log{T_{\rm  eff}}$=  4.40.  The  relative  contribution of  hydrogen
($L_{\rm H}$=$L_{\rm CNO}+L_{\rm pp}$  ) and helium burning to surface
luminosity is  displayed in the lower  panel of Fig.   6. In addition,
the ratio  $L_{\rm pp}/L_{\rm H}$ is  shown.  Some features  of Fig. 6
deserve  comments. To  begin with,  note that  at early  times nuclear
burning via the CN cycle  mostly contributes to the surface luminosity
of the star.   After $7 \times 10^3$ yr of  evolution (that is shortly
after the  remnant reaches the maximum effective  temperature), the CN
cycle reactions abruptly cease, and the model begins to descend to the
white dwarf  domain and the surface  luminosity of the  star begins to
decline steeply. At these stages, helium burning contributes almost as
much to  surface luminosity as hydrogen  burning.  Thereafter, nuclear
burning plays  a minor  role in  the evolution of  the star,  which is
dictated   essentially  by   neutrino  losses   and  the   release  of
gravothermal energy.   The maximum contribution of  nuclear burning to
surface  luminosity  during  the  white  dwarf regime  occurs  in  the
effective temperature range of 35000  to 20000K (at $5 \times 10^6$ to
4 $\times  10^7$ yr)  of evolution.  During  this phase  of evolution,
nuclear burning  is due almost entirely  to CN cycle  reactions at the
base of  the hydrogen layer.  These results  quantitatively agree with
those of  Iben \&  MacDonald (1986)  in the frame  of 0.6  \msun white
dwarf models with  time dependent diffusion and $Z=$  0.02.  At an age
of $\approx 10^8$ yr, the  energy production due to CN cycle reactions
falls below  that due to the  proton-proton chain and by  the time the
white dwarf has  reached the domain of the  ZZ Ceti instability strip,
the   nuclear  energy   production   is  almost   entirely  from   the
proton-proton  chain.  During this  stage,  the  ratio  of nuclear  to
surface luminosity reaches a local  maximum. Note that at the hot edge
of ZZ  Ceti region, nuclear burning  contributes at most 10  \% to the
surface luminosity output, and this reduces  to 5 \% when the red edge
is reached.  Eventually, hydrogen burning becomes virtually extinct at
the lowest luminosities that we compute.

Once  the remnant  has settled  upon its  cooling track,  its chemical
abundance  distribution  will  be  strongly modified  by  the  various
diffusion   processes  acting  during   white  dwarf   evolution.   To
illustrate this important aspect, we plot in Fig.  7 the abundances by
mass of $^1$H, $^3$He, $^4$He,  $^{12}$C and $^{14}$N as a function of
the outer  mass fraction $q$  ($q= 1- M_r/M_*$,  so the centre  of the
star  corresponds to $\log  q=0$) at  various epochs  characterized by
values  of $\log{L/\rm  L_{\odot}}$ and  $\log{T_{\rm  eff}}$ (numbers
given  in  parentheses).   In  addition, the  nuclear  energy  release
$\epsilon_{\rm nuc}$ (in erg/gr/s)  due to hydrogen and helium burning
shells  are  shown  as  thin  lines.   Panel  7a  shows  the  chemical
stratification before the star  reaches the point of maximum effective
temperature at high luminosities.  In the outermost layers, abundances
correspond essentially  to those assumed for  the interstellar medium.
The deeper layers in the helium  buffer zone show a CNO abundance that
is different  from the interstellar  one, because hydrogen  burning in
earlier  evolutionary  phases processed  essentially  all the  initial
$^{12}$C into $^{14}$N.  Panel 7b depicts the situation somewhat later
when the star is at the start of the cooling branch $5 \times 10^5$ yr
after mass  loss stops. During these evolutionary  stages diffusion is
barely noticeable except for  the outermost layers where gravitational
settling causes hydrogen to float  to the surface and heavier elements
to  sink down.  During  this part  of evolution,  chemical composition
changes primarily as  a result of nuclear burning via  the CN cycle at
the base  of the hydrogen  envelope.  With further cooling  (panels 7c
and 7d),  the action of gravitational settling  and chemical diffusion
is apparent.  The effect of  chemical diffusion is clearly seen at the
chemical  interfaces  where  large  abundance gradients  exist.  As  a
result,  in the  helium buffer  there is  a tail  of hydrogen  and, at
advanced  stages,  a tail  of  carbon from  the  bottom.  The tail  of
hydrogen chemically diffusing inwards to hotter layers (thus favouring
the occurrence  of hydrogen burning, see below)  is clearly noticeable
in  these figures,  as well  as  the gravitational  settling of  heavy
elements from the outer  layers.  The chemical stratification when the
model reaches the domain of the ZZ Ceti after $3 \times 10^8$ years is
displayed  in panel  7e.  Needless  to say,  diffusion  processes have
substantially altered  the chemical abundance  distribution within the
star as compared  with the initial stratification at  the start of the
cooling branch.   In particular, the  star is characterized by  a pure
hydrogen envelope of mass $3.5  \times 10^{-5}$ \msun plus a wide tail
of  hydrogen  reaching very  deep  layers  (by  this time,  the  total
hydrogen  content left  in  the  star amounts  to  $7 \times  10^{-5}$
\msun).  Indeed,  the tail of hydrogen distribution  reaches a maximum
depth by this  epoch. Note also the tail of  carbon diffusing into the
helium buffer from deeper layers.  With further cooling, the diffusive
tail  of  hydrogen  begins  to  retreat outwards  again,  because  the
increasing  electron degeneracy  causes chemical  diffusion  to became
less important  and the inward  diffusion of hydrogen is  stopped. The
chemical profile corresponding to the  last computed model is shown in
panel 7f. It  is clear that, except for the  outermost layers at large
ages, the situation of diffusive equilibrium has not been reached even
at  the latest  evolutionary stages  we computed.   So,  the diffusive
equilibrium approach  used to infer  the chemical profile of  DA white
dwarfs  in some  pulsation  studies  of these  stars  is not  entirely
appropriate.    Another   interesting   observation   concerning   the
importance of diffusion in the  element distribution is related to the
evolution  of $^{12}$C and  $^{14}$N abundances  in the  buffer layer.
Indeed,  initially  the  $^{14}$N   abundance  in  the  helium  buffer
overwhelms that of  $^{12}$C (see panel 7a), but  at the latest stages
computed, $^{14}$N  is less abundant  than $^{12}$C in almost  all the
stellar interior,  despite nuclear burning  has processed considerable
$^{12}$C into $^{14}$N during white dwarf evolution.  The same finding
has  been reported  by Iben  \&  MacDonald (1986),  although in  their
calculations, the $^{12}$C abundance exceeds the $^{14}$N abundance by
the time  the ZZ Ceti  domain is reached.  This is earlier  than takes
place in  our calculation.  In  part, this difference between  the two
sets  of calculations  can  be understood  on  the basis  that in  the
present study we have  included thermal diffusion, whilst this process
has  been  neglected by  Iben  \&  MacDonald  (1986). Because  thermal
diffusion acts in the  same direction that gravitational settling, the
neglect  of the former  leads to  a larger  $^{12}$C abundance  in the
helium  buffer.   In this  context,  this  effect  is expected  to  be
responsible  for  differences in  the  nuclear  burning luminosity  as
compared with the situation  in which thermal diffusion is considered.
To verify these assertions, we have computed the white dwarf evolution
of the remnant but without considering thermal diffusion.  In fact, we
found that in such a case the resulting $^{12}$C abundance by the time
the ZZ Ceti  domain is reached becomes somewhat larger  at the tail of
its distribution; in addition,  the contribution of nuclear burning to
surface luminosity increases (by at most 20 \%) when thermal diffusion
is not taken into account.

An aspect  worthy of  discussion is the  effect that  chemical element
diffusion  has on  nuclear  burning.  This  important  point has  been
studied in  detail by Iben \&  MacDonald (1985) in the  case of models
with $Z$=0.001. The situation for  the case $Z$= 0.02 is qualitatively
similar though  some differences  exist.  In broad  outline, diffusion
causes CN cycle reactions to be  efficient for a longer period of time
than  when  diffusion  is   neglected.   This  can  be  understood  by
inspecting  the distribution  of nuclear  energy generated  within the
star  given in  Fig.  7. Indeed,  as  a result  of hydrogen  diffusing
downwards to hotter layers into the helium buffer and carbon diffusing
upwards  from  the carbon-rich  zone  through  the  buffer layer,  the
production  of   nuclear  energy   via  CN  cycle   reactions  remains
significant  for   a  long   period  of  time   in  the   white  dwarf
evolution.  The location of  the peak  in $\epsilon_{_{\rm  CN}}$ gets
deeper as  evolution proceeds, reaching a  maximum depth at  a mass of
$\approx 1.7  \times 10^{-3}$ \msun  below the stellar  surface.  Note
that despite  chemical diffusion leading to  appreciable abundances of
$^1$H and $^{12}$C  in the helium buffer, this has  not resulted in an
enhanced  nuclear  energy  release.   However, this  could  have  been
radically different had the helium  buffer been less massive than ours
( $\approx$  0.003 \msun).  In  fact, Iben \& MacDonald  (1986) showed
that a  hydrogen shell  flash is initiated  when the star  has already
settled upon the  white dwarf cooling track if  the helium buffer mass
is  as small as  0.001 \msun,  and that  the thermonuclear  runaway is
avoided if it is as massive as 0.005 \msun.  As mentioned earlier, how
massive the helium buffer can be  depends (for a progenitor of a given
initial stellar mass)  on the precise phase in  the helium shell flash
cycle at which the star departs  from the AGB.  So, the role played by
the CN  cycle reactions in  producing nuclear energy during  the white
dwarf stage  will also depend on  the phase of the  helium shell flash
cycle at which  departure from the AGB occurs.   Finally, we note that
as  cooling  proceeds, the  mass  range  over  which hydrogen  nuclear
burning  extends  becomes  wider  as  fast as  the  tail  of  hydrogen
distribution penetrates inwards.  As can be seen, the maximum width is
found at the ZZ Ceti region,  so in principle the mode stability in DA
white  dwarf models  could be  affected by  the occurrence  of nuclear
burning  over a  considerable mass  range. In  this connection,  it is
worth commenting on the fact that shell nuclear burning has been found
responsible    for     instability    of    g-modes     through    the
$\epsilon$-mechanism in models of pre-white dwarfs (see Kawaler et al.
1986 and Kawaler 1988).  However, the periods of the excited modes are
so  short that  they have  not been  observed in  such stars  (Hine \&
Nather  1987). Anyway, we  find that  hydrogen burning  contributes at
most 10 \% to surface luminosity during the ZZ Ceti stage.  We want to
stress that  according to our calculations nuclear  burning during the
white dwarf stage does not represent a major source of energy even for
the   maximum   hydrogen  mass   allowed   by   the  pre-white   dwarf
evolution. This  is a consequence  not only of  the fact that  in this
calculation we assume a high initial metallicity ($Z= 0.02$) (see Iben
\& MacDonald 1986)  but also that, as mentioned,  the inclusion of the
thermal diffusion process lessens the hydrogen burning contribution to
surface luminosity.

Because the shape of the  chemical composition profile is a key factor
in determining the  g-mode periods of DAV white  dwarfs, we compare in
Fig.  8  the  chemical  stratification  of  the  white  dwarf  at  the
instability domain (thick lines)  with that corresponding at the start
of the  cooling branch  (thin lines). The  role of diffusion  is again
clearly  emphasized in this  figure. Notably,  near-discontinuities in
the initial  abundance distribution  are smoothed out  considerably by
element diffusion.  In fact, diffusion processes strongly modifies the
slope of the chemical profiles in  the outer layers on the white dwarf
cooling  track. As  well  known,  such regions  are  critical for  the
pulsation properties of white dwarfs. In the next section, we describe
the effect of  such chemical profiles on the  main quantities entering
adiabatic pulsation equations.

In what follows,  we shall comment on the  implications of the results
we have discussed thus far for those quantities entering the pulsation
equations  that depends  essentially on  the profile  of  the chemical
composition.   Here,  we shall  limit  ourselves  exclusively to  that
aspect, deferring  a thorough discussion  of the consequences  for the
global pulsation properties of the model to a future work.

\section{QUANTITIES FOR ADIABATIC PULSATION ANALYSIS }

Here  we describe  at some  length  the characteristics  of the  basic
variables  which are  relevant for  adiabatic pulsation  analysis.  We
shall  concentrate particularly  on the  Brunt-V\"ais\"al\"a frequency
(hereafter BVF) and  the Ledoux term (B). To  perform our analysis, we
pick out  a model within the  ZZ Ceti instability  strip ($T_{\rm eff}
\sim 12000$ K).

Roughly speaking, the BVF is  the oscillation frequency of a parcel of
stratified stellar  fluid when  it is vertically  (radially) displaced
from its equilibrium level, and buoyancy acts as restoring force.  The
BVF (N) is defined as (Unno et al. 1989):

\begin{equation} \label{eq1}
N^2  \equiv  g\  \left(\frac{1}{\Gamma_1}\
\frac{d\ln P}{dr}  - \frac{d\ln \rho}{dr}\right).
\end{equation}

As well  known, the shape of  the BVF is directly  responsible for the
global  characteristics  of the  period  spectrum  in pulsating  white
dwarfs.  It  can be demonstrated  by employing a local  analysis (see,
e.g.   Unno   et  al.   1989)  that  the   condition  for  propagating
(non-evanescent) g-modes is  that the frequency squared, $\sigma^{2}$,
must be less than both $L_{\ell}^{2}$ and $N^{2}$, where $L_{\ell}$ is
the  (acoustic)  Lamb  frequency  (the  other  critical  frequency  of
non-radial stellar oscillations).  The  region of the model where this
condition is  accomplished is  the propagation zone  of the  mode (see
propagation diagrams  in e.g.   Cox 1980 and  Unno et al.   1989).  In
particular for white dwarfs, the BVF reaches very small values deep in
the degenerate core, thus excluding the possibility for propagation of
low order g-modes (short  periods).  Instead, these modes propagate in
the  envelope of the  star, and  are therefore  very sensitive  to the
detailed structure of the outer regions.

From the computational point of view,  the treatment of the BVF in the
interior of  white dwarf  stars has been  thoroughly discussed  in the
past  (see Tassoul et  al.  1990,  and in  particular Brassard  et al.
1991).  These studies illustrate the numerical problems and systematic
errors that result from the computation of the BVF when it is assessed
directly  from  its  definition  (Eq.  \ref{eq1}).   Brassard  et  al.
(1991) have  shown that,  in the  frame of  their stellar  models, the
profile  of  BVF  obtained  from  Eq.   (\ref{eq1})  can  lead  to  an
unreliable pattern of g-mode periods.  As pointed out in that article,
the reason for that is  not the computing of the numerical derivatives
themselves,   but  instead,   that  the   direct  employment   of  Eq.
(\ref{eq1})  for strongly  degenerate  matter typical  of white  dwarf
interiors  implies subtracting  two large  quantities that  are nearly
equal, which produces, besides spurious structures in the $N$ profile,
a global shift  towards greater BFV values, particularly  in the inner
core (see this notable effect in  Fig.  10a of Brassard et al.  1991).
This leads to  an increase in the eigenfrequencies  of g-modes, or, in
other words, to a  displacement towards shorter periods.  In addition,
the region of period formation,  as sketched from the weight functions
(see Brassard et al. 1991), is strongly affected.

To  overcome these  difficulties, Brassard  et al.   (1991)  provide a
computational  strategy appropriate for  degenerate objects,  known in
the literature  as ``modified Ledoux'' treatment.   According to these
authors, $N^2$ in white dwarf models is to be computed as:

\begin{equation} \label{eq2}
N^2   =   \frac{g^2\   \rho}{P}\   \frac{\chi_{_{\rm   T}}}{\chi_{\rho}}\
\left(\nabla_{\rm ad} - \nabla + B \right),
\end{equation}

\noindent  where  $\chi_{_{\rm T}}$  ($\chi_{\rm  \rho}$) denotes  the
partial logarithmic pressure derivative  with respect to $T$ ($\rho$),
$\nabla$  and   $\nabla_{\rm  ad}$   are  the  actual   and  adiabatic
temperature  gradients, respectively,  and  $B$, the  Ledoux term,  is
given by

\begin{equation} \label{eq3}
B=-\frac{1}{\chi_{_{\rm T}}} \sum^{n-1}_{{\rm i}=1} \chi_{_{X_{\rm i}}}
\frac{d\ln {X}_{\rm i}}{d\ln P}.
\end{equation}

Here $X_{\rm  i}$ is the abundance by  mass of specie $i$,  $n$ is the
total number of considered species and

\begin{equation} \label{eq4}
\chi_{_{X_{\rm i}}}= \left( \frac{\partial \ln{P}}
{\partial \ln{X_{\rm i}}} \right)_{\rho,T,\{X_{\rm j \neq i}\} }.
\end{equation}

This formulation has the  advantage of avoiding the problems mentioned
above,  and  at   the  same  time  it  explicitly   accounts  for  the
contribution to $N^2$  from any change in composition  in the interior
of model  (the zones  of chemical transition)  by means of  the Ledoux
term (Eq.  \ref{eq3}).  Brassard  et al.  (1992a) stress the relevancy
of  a correct  treatment  of the  BVF  in the  interfaces of  chemical
composition  in stratified  white dwarfs,  particularly  in connection
with the  resonance effect of  modes known as ``mode  trapping''.  The
modified Ledoux treatment is employed in most of the pulsation studies
in white dwarfs \footnote{Except,  e.g., in Gautschy et al.  (1996).}.
Our pulsation code (C\'orsico \& Benvenuto 2001) is also based on such
a formulation.

The Ledoux term  $B$ is an important ingredient  in the computation of
$N^2$.  In most of existing studies, the shape of $B$ is computed from
chemical profiles treated on the basis of diffusive equilibrium in the
trace element approximation (see,  e.g., Tassoul et al. 1990; Brassard
et al.  1991,  1992ab).  The behaviour of $B$  is responsible (through
$N^2$)  for   macroscopic  effects  on  the   period  distribution  in
stratified  white   dwarfs,  such  as  the  mode   trapping  and  mode
confinement (in  the terminology  of Brassard et  al. 1992b).   In the
upper panel of Fig.  9 we show the profiles of abundances for the most
relevant chemical species of our  models (see section Section 3.2). In
the  middle panel we  depict the  corresponding $B$  term, and  in the
bottom panel we plot $N^2$.  The chemical profiles as predicted by our
calculations are  very smooth  in the interfaces,  which give  rise to
extended tails  in the  shape of  $B$.  Also, note  that our  model is
characterized by a chemical interface  in which three ionic species in
appreciable  abundances  coexists: oxygen,  carbon  and helium.   This
transition gives two  contributions to $B$, one of  them of relatively
great magnitude,  placed at  $\log q \sim  -1.4$, and the  other, more
external and  of very low height  at $\log q \sim  -2.2$. This feature
makes a difference as compared  with the results of other authors (see
Tassoul et  al.  1990; Brassard  et al.  1991; 1992ab;  Bradley 1996).
As a last remark, we note  that the contribution of the He/H interface
to $B$ is less than that corresponding to the O/C/He transition.  From
the bottom panel of Fig.  9  we can note each feature of $B$ reflected
in the  shape of the  BVF.  The contributions  of the Ledoux  term are
translated  into extended  bumps  on $N^2$.   Note  the global  smooth
behaviour  of these quantities  when account  is made  of evolutionary
models  with time dependent  element diffusion.   In the  interests of
comparison,  we have  also computed  $B$  and $N^2$  according to  the
prediction  of   the  diffusive  equilibrium  in   the  trace  element
approximation  as  given by  Tassoul  et  al.  (1990).  The  resulting
chemical profile  at the hydrogen-helium  transition as given  by this
approximation is depicted with a thin  lines in the inset of the upper
panel of  Fig. 9.  In agreement  with previous studies,  note that the
equilibrium diffusion approximation leads  to a pronounced peak in the
Ledoux term at that chemical  interface, which translates into a sharp
peak of $N^2$ in that region.

For the  sake of  completeness, we have  computed $N^2$ also  from Eq.
(\ref{eq1}).    After  some   minor  manipulations,   and   using  the
hydrostatic   equilibrium  condition   ($dP/dr  =   -\rho\   g$),  Eq.
(\ref{eq1}) can be rewritten as:

\begin{equation} \label{eq5}
N^2  = - \frac{g^2 \rho}{P}  \left(\frac{d\ln \rho}{d\ln P} -
\frac{1}{\Gamma_1}\right).
\end{equation}

The  derivative $d\ln  \rho /  d\ln P$  in Eq.   (\ref{eq5})  has been
computed  numerically employing  an  appropriate interpolation  scheme
(Benvenuto \& Brunini 2001). This scheme provides the first derivative
for the  interpolated points.  The result  for $N^2$ is  shown in Fig.
10a in terms  of $r/R_{*}$.  In the interests  of comparison, Fig. 10b
shows  $N^2$  computed  using  the  modified  Ledoux  treatment,  from
Eq.(\ref{eq2}).  Note that our  strategy for the numerical derivatives
gives excellent  results, reproducing rigorously to  the finest detail
of  the $N^2$  profile as  computed from  Eq.(\ref{eq2}).   Very small
structures  (numerical  noise)  towards   the  centre  can  barely  be
observed.  Clearly,  in  contrast  to  the  Brassard  et  al.   (1991)
assertions,  our  calculations  based  on  an  appropriate  scheme  of
interpolating, provides  a reliable profile  for $N^2$ in  our stellar
models (see also a similar finding in Gautschy et al. 1996).

\section{CONCLUSION} \label{sec:conclusion}

In this  work we  present new evolutionary  calculations for  DA white
dwarf stars.  The calculations  fully take into account time dependent
element diffusion, nuclear burning and  the history of the white dwarf
progenitor in  a self-consistent way.  The primary  application of our
evolutionary  models  will  be  the  exploration  of  their  pulsation
properties in future papers. Evolutionary calculations are carried out
by means of  a detailed and up-to-date evolutionary  code developed by
us at  La Plata  Observatory. The code  has been employed  in previous
studies   of  white  dwarf   evolution  and   it  has   recently  been
substantially modified  to study  the evolutionary stages  previous to
the  formation of  white dwarfs.   Briefly, up-to-date  OPAL radiative
opacities for different  metallicities, conductive opacities, neutrino
emission rates and  equation of state are considered.  In addition, we
include a network of 30  thermonuclear reaction rates for hydrogen and
helium burning and the evolution of 13 chemical species resulting from
nuclear burning is followed via an implicit method of integration. Our
code  enables  us  also  to  compute the  evolution  of  the  chemical
abundance distribution due to the processes of gravitational settling,
thermal  and chemical  diffusion.   The treatment  for time  dependent
diffusion  is based  on  the formulation  of  multicomponent gases  by
Burgers (1969).

Specifically, we  follow the evolution  of an initially 3  \msun model
from the ZAMS  (the adopted metallicity is 0.02) all  the way from the
stages of hydrogen and helium burning  in the core up to the thermally
pulsing  phase at  the tip  of the  AGB.  By  the end  of  core helium
burning,  the star  experiences  a  total of  40  micro-pulses of  low
surface luminosity  amplitude.  Afterwards, evolution  proceeds to the
phase of  major thermal pulses on  the AGB, during  which helium shell
burning becomes unstable (helium shell flashes). After experiencing 11
thermal pulses, the model is  forced to evolve towards its white dwarf
configuration  by invoking  strong  mass loss  episodes. Evolution  is
pursued  till the  domain of  the  ZZ Ceti  stars on  the white  dwarf
cooling branch.

We find that if departure from the AGB occurs early in the inter-flash
cycle during  stationary helium burning,  then, after the end  of mass
loss episodes (at log $T_{\rm eff}$= 3.8), the star returns to the red
where, in the  meantime, the hydrogen mass is  considerably reduced by
hydrogen shell  burning over a  long period of  time. As a  result, no
planetary nebula  is produced.  Indeed, the remnant  needs about 48000
yr  to reach  an  effective  temperature of  30000K  required for  the
excitation of  the nebula.   This is in  agreement with  previous full
evolutionary calculations of post-AGB  stars by Mazzitelli \& D'Antona
(1986) and Wood \& Faulkner (1986).  The mass of hydrogen that is left
at the start  of the white dwarf cooling branch is  about $ 1.5 \times
10^{-4}$ \msun, and this is reduced to $7 \times 10^{-5}$ \msun due to
the interplay of further nuclear  burning and element diffusion by the
time  the ZZ  Ceti domain  is reached.   Because we  have  not invoked
additional  mass loss episodes  during the  planetary nebula  stage or
early  during the  cooling  branch,  the quoted  value  for the  final
hydrogen mass should be considered  as an upper limit. Another feature
of interest shown by our results  is related to the mass of the helium
buffer. We  find that when the  progenitor leaves the  AGB, the helium
buffer mass is  $ \approx 3 \times 10^{-4}$ \msun,  but as the remnant
returns to the  red, the buffer mass increases  to 0.0024 \msun, which
is massive enough to  prevent a diffusion-induced hydrogen shell flash
from occurring  on the white  dwarf cooling track.  Thus,  we conclude
that if departure from the AGB takes place during the quiescent helium
burning  phase, a  self-induced  nova  event as  proposed  by Iben  \&
MacDonald (1986) is not possible.

In  agreement with  Iben  \&  MacDonald (1985)  we  find that  element
diffusion  strongly modifies the  distribution of  chemical abundances
during the white dwarf  cooling. Near discontinuities in the abundance
distribution  at the  start  of the  cooling  branch are  considerably
smoothed out by the diffusion processes by the time the ZZ Ceti domain
is reached. Our calculations also show that the situation of diffusive
equilibrium has not yet been reached (except for the outermost layers)
by  the time  the ZZ  Ceti  domain is  reached.  Indeed,  there is  an
appreciable  evolution of the  chemical abundances  at such  stages of
evolution. With  regard to nuclear burning,  we find that  it does not
represent a  major source of energy  during the white  dwarf stage, as
expected for a progenitor star of initially high metallicity.  We also
find  that thermal diffusion  lessens even  further the  importance of
nuclear burning.

Finally,  we have  discussed at  some length  the implications  of our
evolutionary  models for  the main  quantities relevant  for adiabatic
pulsation  analysis.  We find  that the  shape of  the Ledoux  term is
markedly  different from that  found in  previous detailed  studies of
white dwarf pulsations. This is due partly to the effect of smoothness
in the chemical distribution  caused by element diffusion, which gives
rise  to   less  pronounced   peaks  in  the   Ledoux  term   and  the
Brunt-V\"ais\"al\"a frequency.
             
We believe that the evolutionary models presented in this work deserve
attention  concerning their pulsation  properties and  mode stability.
We expect  the pulsation properties  of our models to  show noticeable
differences as  compared with  those encountered in  previous studies.
Last but  not least,  fundamental aspects related  to the  analysis of
mode trapping in white dwarfs would be worth carrying out in the frame
of  the  present  models.   In   this  sense  we  speculate  that  the
oscillation  kinetic  energy  of  some  modes  could  be  considerably
affected as a result of the shape of the Brunt-V\"ais\"al\"a frequency
as  predicted by  our  evolutionary models.   The  assessment of  such
further  aspects of  white  dwarf stars  is  beyond the  scope of  the
present paper.

Before closing, we  would like to comment on the  well known fact that
there are still many uncertainties  in the theory of stellar evolution
that  prevent us  from being  completely confident  about  the initial
chemical  stratification  of white  dwarf  stars.   For instance,  the
degree  to  which semi-convection  and  overshooting  affect the  core
chemical stratification is not known from first principles and in most
of studies  in which this aspect  is addressed, they  are treated very
roughly.   In this  study  overshooting and  semi-convection were  not
included.   So,  in  principle,  the  profile  of  the  core  chemical
composition  could be  somewhat different  according to  whether these
processes are  taken into account  or not.  In addition,  the envelope
chemical stratification depends quite  sensitively on the phase of the
helium shell flash cycle at  which the progenitor star leaves the AGB.
This  is particularly  true regarding  the mass  of the  helium buffer
layer. Finally,  mass loss episodes during the  planetary nebula stage
are  expected to  reduce the  hydrogen envelope  mass of  the post-AGB
remnant.   We  plan  to  explore   some  of  these  issues  and  their
consequences for pulsation analysis in a future work.

Evolutionary models presented in this paper are available upon request
to the authors at their e-mail addresses.

AMS warmly  acknowledges Arnold Boothroyd for making  available to him
the latest set of routines  for computing OPAL radiative opacities for
different metallicities. It  is a pleasure to thank  our referee, Paul
Bradley, whose comments and  suggestions improved the original version
of this work.

\newpage


\begin{figure*}
\epsfysize=600pt
\epsfbox{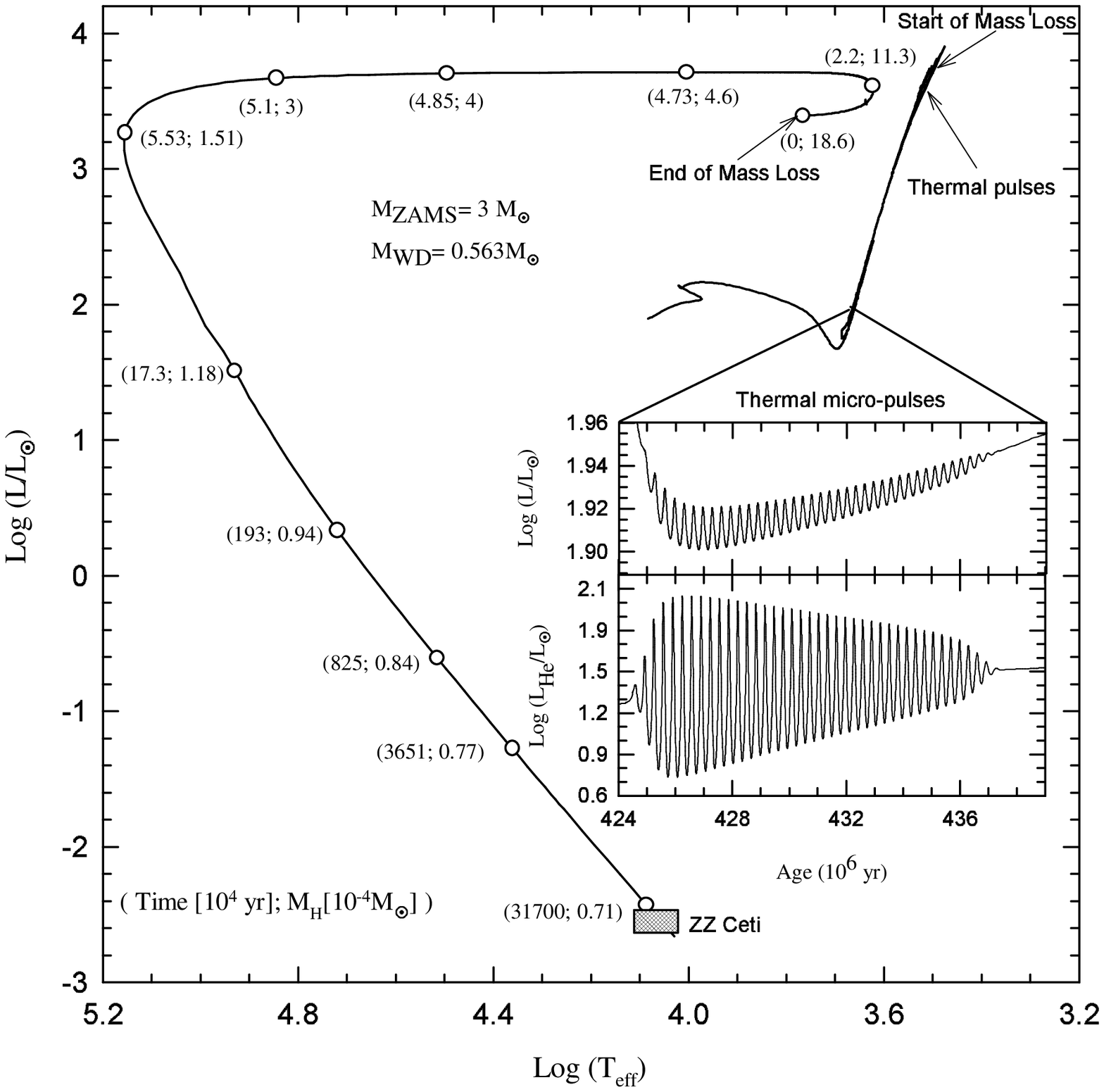} 
\caption{Hertzsprung - Russell diagram for the evolution
of our 3  M$_{\odot}$ stellar model, from the ZAMS to the white dwarf
stage.  For clarity, the evolutionary stages corresponding to the mass
loss  phase are  not shown.   Numbers in  parenthesis  besides circles
along the track  give the age (in $10^4$ yr) measured  from the end of
mass  loss and  the mass  of hydrogen  in the outer  layers in units of
$10^{-4}$ M$_{\odot}$. The domain of  the ZZ Ceti instability strip is
shown as a shaded region.  As  a result  of mass  loss episodes,  the  
stellar mass
decreases from  3 to  0.563 M$_{\odot}$.  Note  that after the  end of
mass loss, the star returns towards lower effective temperatures where
it burns an appreciable fraction of its hydrogen content.  During this
phase,  evolution  proceeds  very  slowly.   The  inset  displays  the
evolution  of surface  (top panel)  and helium-burning  (bottom panel)
luminosities expressed in solar  units during the thermal micro-pulses
phase towards the end of helium burning in the stellar core.} 
\end{figure*}

\begin{figure*}
\vskip 2.5cm
\epsfxsize=400pt
\epsfbox{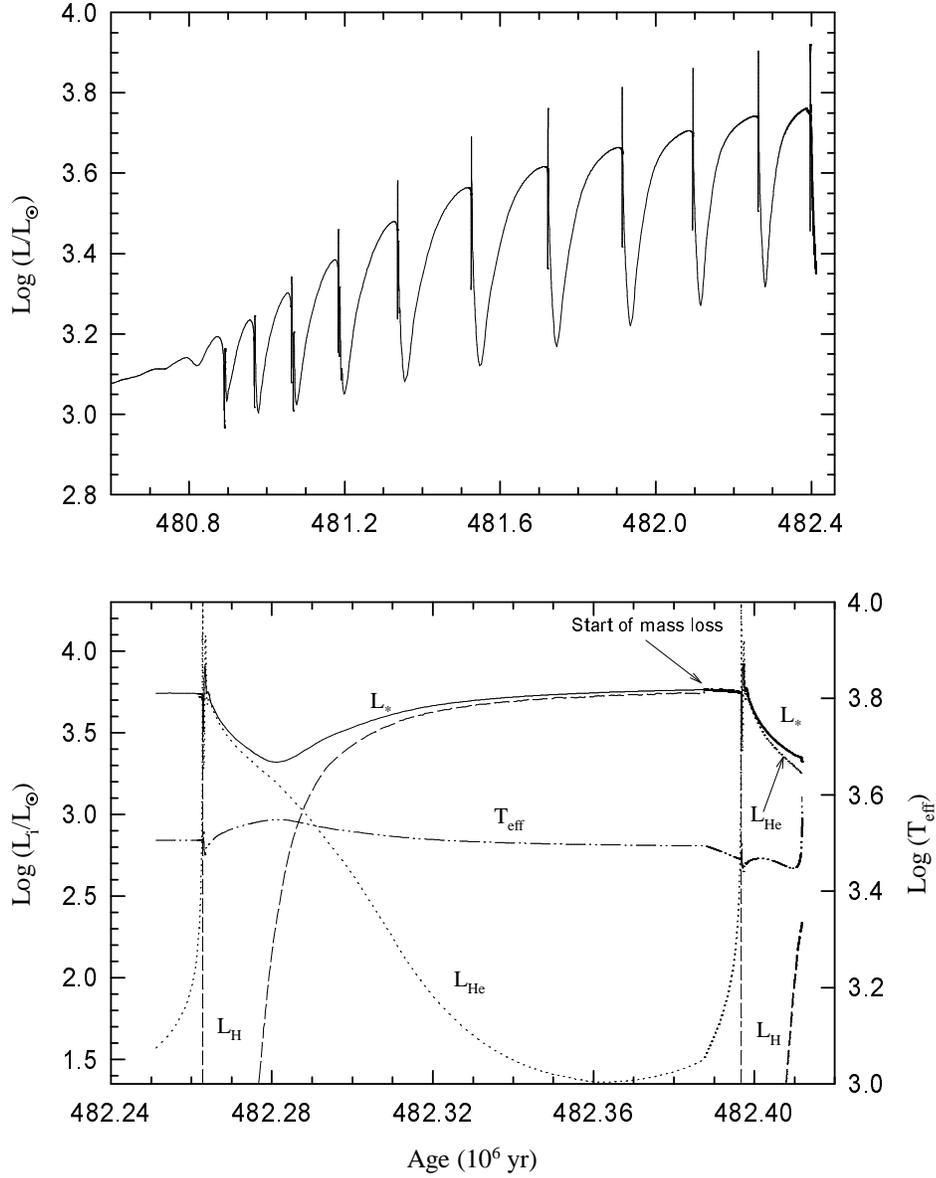} 
\caption{Top panel:  Evolution of surface luminosity  (in solar units)
during the  phase corresponding to the  helium thermal pulses  for a 3
M$_{\odot}$  model at the  top of  the AGB.   The model  experiences a
total of 11  pulses before departing from the AGB as  a result of mass
loss episodes. The time scale is given in million years from the ZAMS.
Bottom panel: Evolution  of surface luminosity ($L_*$), helium-burning
luminosity ($L_{\rm  He}$), hydrogen-burning luminosity  ($L_{\rm H}$)
and effective temperature ($T_{\rm  eff}$) for the 3 M$_{\odot}$ model
during and between  its tenth and eleventh thermal  pulses.  Note that
the model  departs from  the AGB shortly  after the occurrence  of the
eleventh pulse peak when helium shell burning is dominant.}
\end{figure*}

\begin{figure*}
\vskip 4.0cm
\epsfysize=500pt
\epsfbox{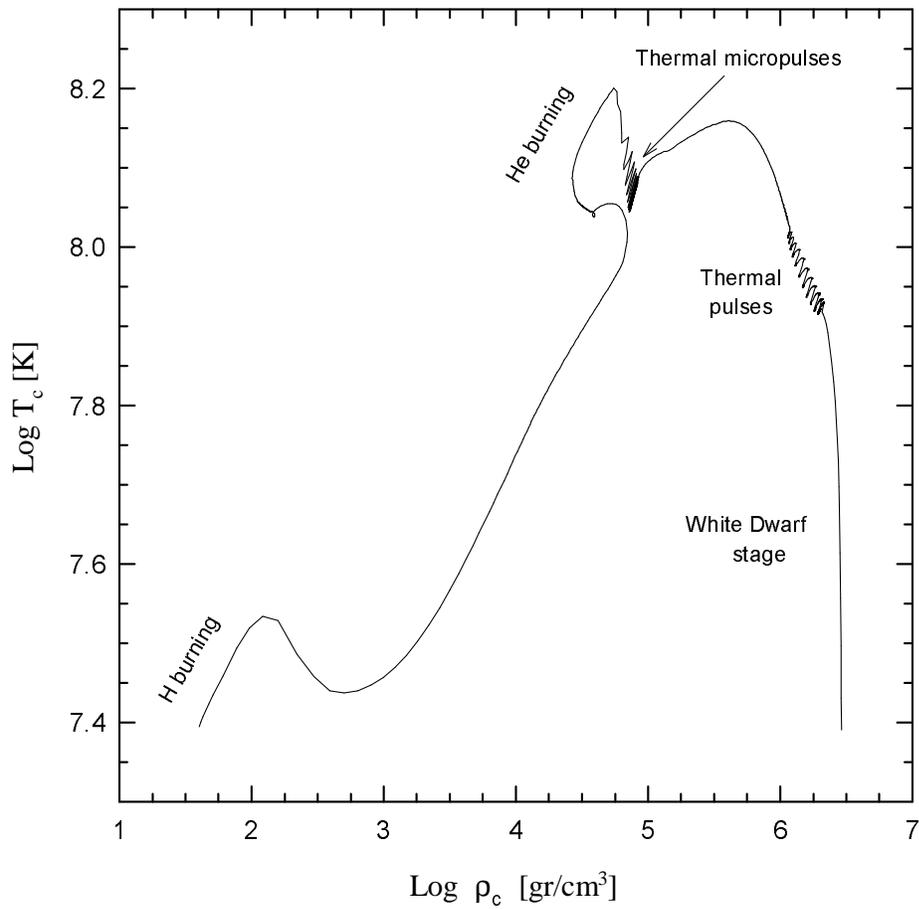} 
\caption{Central temperature  versus central density  corresponding to
the evolution  of the 3  M$_{\odot}$ model all  the way from  the ZAMS
through the AGB  towards the stage of white  dwarf.  Relevant episodes
during  the life  of the  star such  as hydrogen  and  helium burning,
thermal micro-pulses and pulses are indicated as well.}
\end{figure*}

\begin{figure*}
\vskip 3.25cm
\epsfysize=400pt
\epsfbox{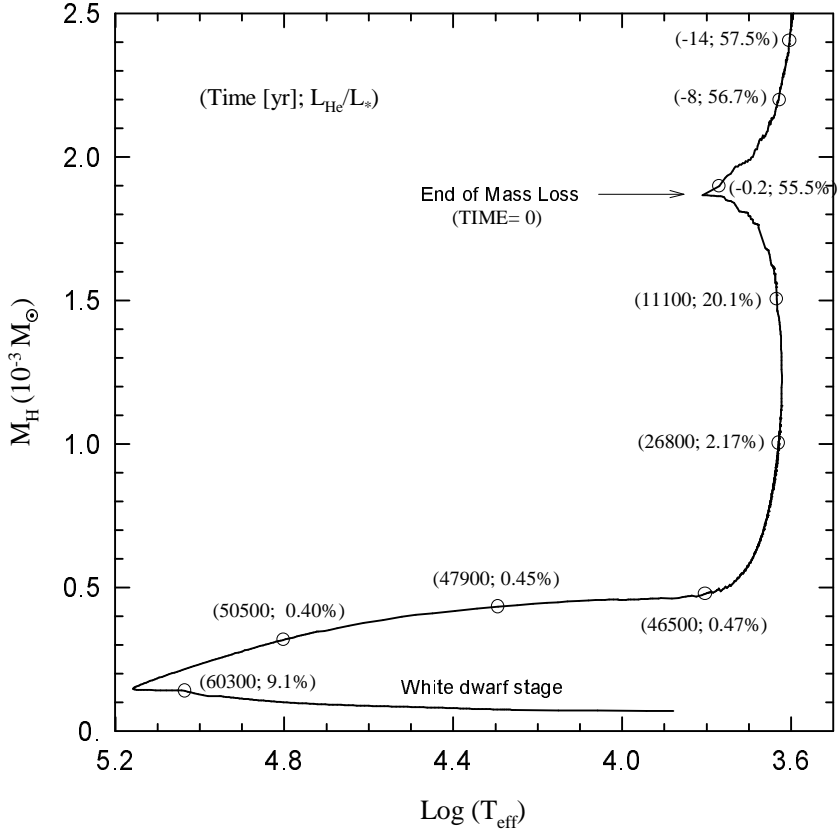} 
\caption{The mass  of hydrogen in  outer layers in units  of $10^{-3}$
solar masses  as a function of effective  temperature for evolutionary
stages following and immediately preceding  the end of mass loss.  The
results correspond to a stellar remnant of $\approx$ 0.563 M$_{\odot}$
resulting  from  the  evolution  of  a  main  sequence  progenitor  of
initially  3 M$_{\odot}$, which  departed the  AGB while  helium shell
burning dominates.   Numbers in parentheses  along the curve  give the
age (in  years) counted from the  moment that mass  loss is suppressed
and  the  percent  contribution  of  helium  burning  to  the  surface
luminosity.   Note that  when mass  loss is  halted, the  star returns
towards lower  effective temperatures where,  over a time  interval of
$\approx$  40000 yr,  nuclear  burning reduces  the  hydrogen mass  by
almost  a factor  of  4.  Afterwards,  the  helium burning  luminosity
becomes  negligible  and the  remnant  resumes  its evolution  towards
higher  effective temperatures,  eventually reaching  the  white dwarf
stage  with  a  hydrogen  content  of  $\approx  1.5  \times  10^{-4}$
M$_{\odot}$.   Subsequent   hydrogen  burning  reduces   the  residual
hydrogen envelope  mass by an  additional factor of 2  before reaching
the domain of the ZZ ceti stars.}
\end{figure*}

\begin{figure*}
\vskip 4cm
\epsfysize=500pt
\epsfbox{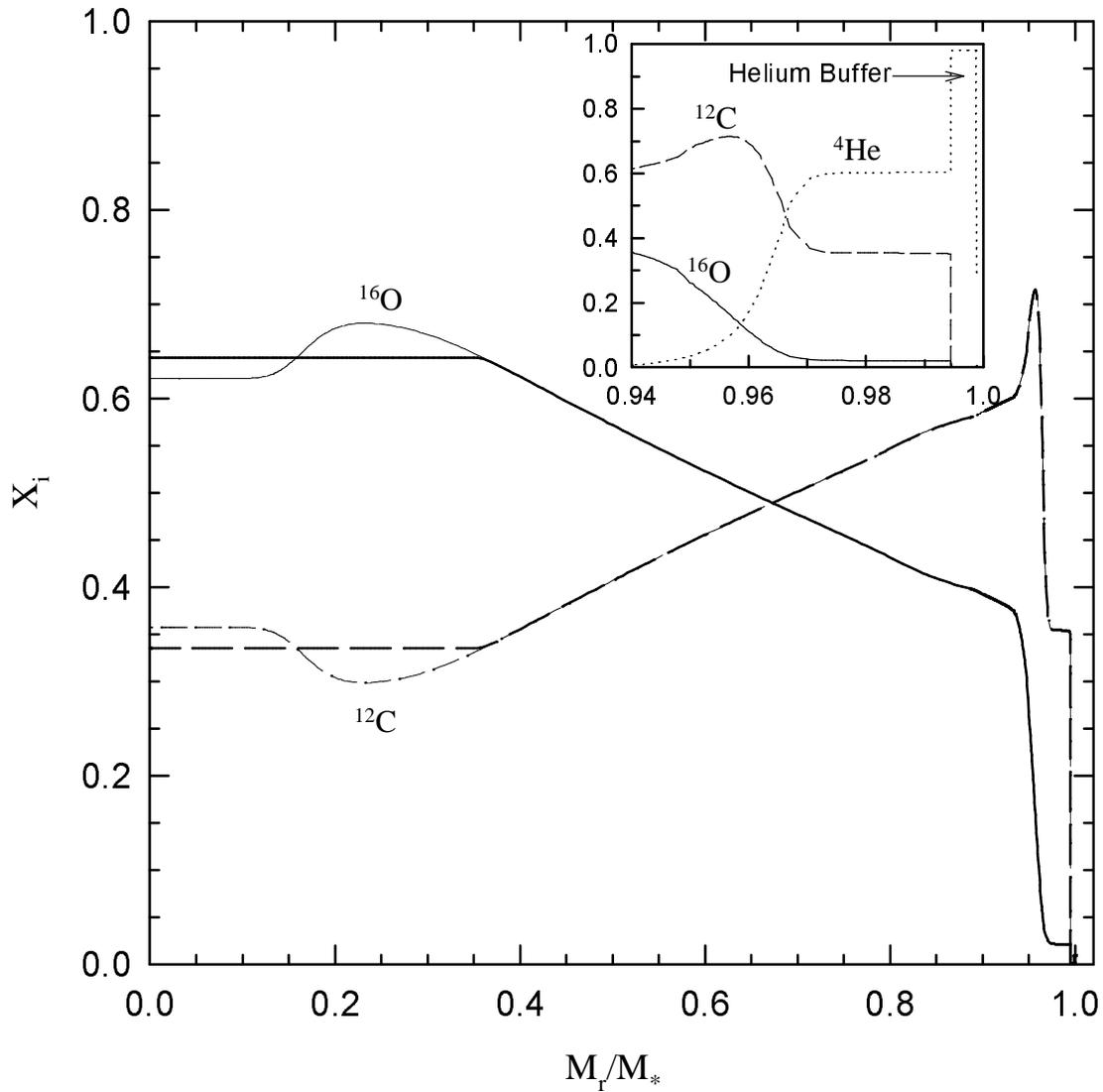} 
\caption{Carbon and oxygen abundance profiles in the core of the 0.563
M$_{\odot}$  remnant for  a  progenitor with  initially 3  M$_{\odot}$
shortly after  the end of mass  loss episodes. Thick  lines depict the
profile  after the  chemical rehomogenization  due  to Rayleigh-Taylor
instability has  occurred in  the central region  of the  star, whilst
thin lines correspond to the situation before the occurrence of such a
rehomogenization.   The  inset shows  the  abundance distribution  for
carbon, oxygen and helium in the outer layers.}
\end{figure*}

\begin{figure*}
\vskip 3cm
\epsfysize=500pt
\epsfbox{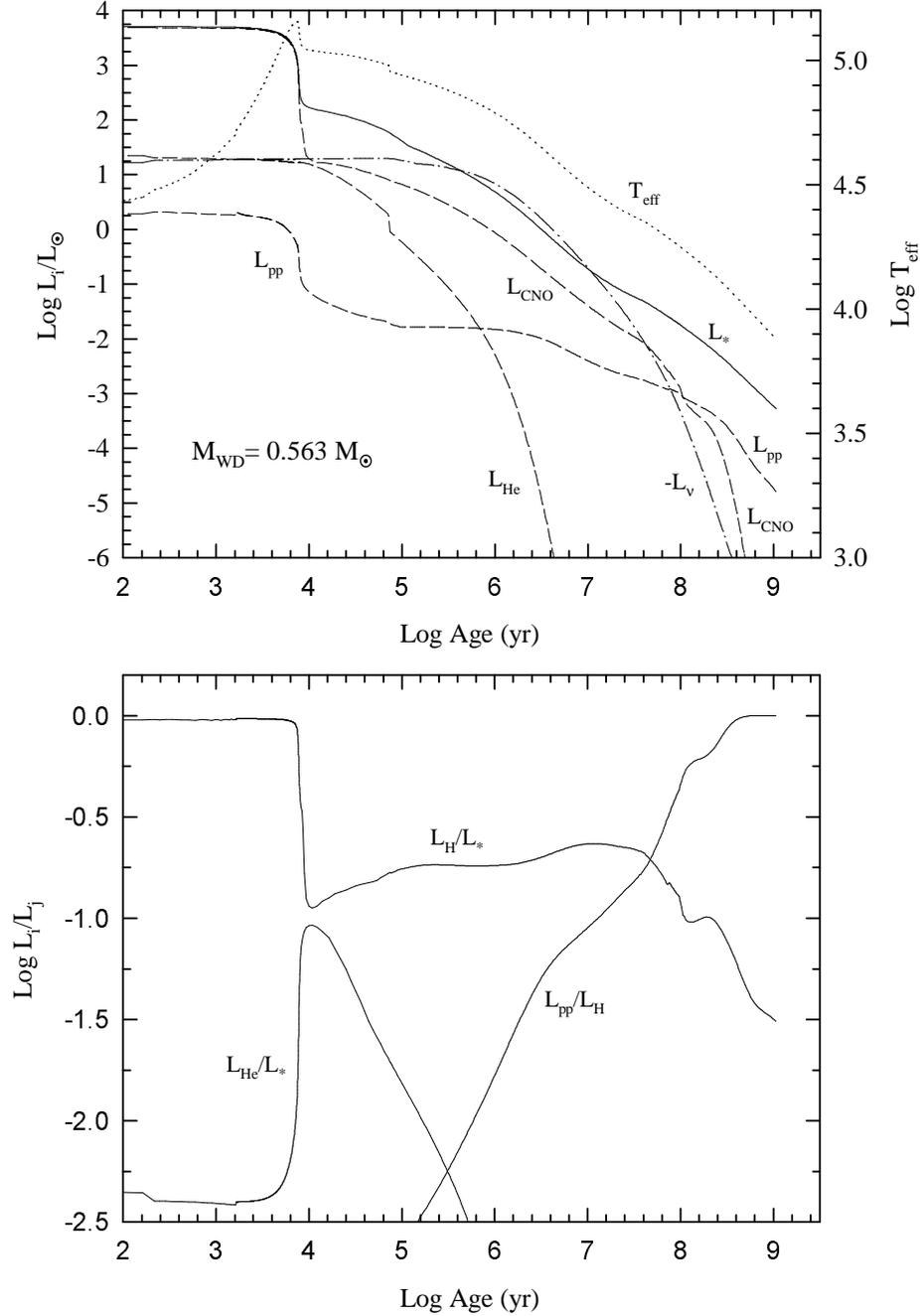} 
\caption{The  top panel plots  different luminosity  contributions (in
solar units)  as a  function of time  for the 0.563  M$_{\odot}$ white
dwarf   remnant:  surface   luminosity,  $L_*$,   luminosity   due  to
proton-proton  reactions, $L_{\rm pp}$,  CNO bi-cycle,  $L_{\rm CNO}$,
helium  burning, $L_{\rm  He}$, and  neutrino losses,  $L_{\nu}$.  The
evolution  of the effective  temperature is  also plotted.   Here, the
zero-age point  corresponds to the  moment when the model  reaches the
point  defined  by  $\log{L/\rm  L_{\odot}}$=  3.71  and  $\log{T_{\rm
eff}}$= 4.40.   The bottom panel displays  luminosity ratios.  $L_{\rm
H}$ stands  for the complete  luminosity as given by  hydrogen burning
($L_{\rm H}= L_{\rm  CNO}+L_{\rm pp}$).  Note that by  the time the ZZ
Ceti domain is reached, hydrogen burning, which is almost entirely due
to  proton-proton  reactions, attains  a  local  maximum, providing  a
contribution to the surface luminosity output of about 10 \%.  Time is
in years.}
\end{figure*}

\begin{figure*}
\vskip 3.5cm
\epsfysize=500pt
\epsfbox{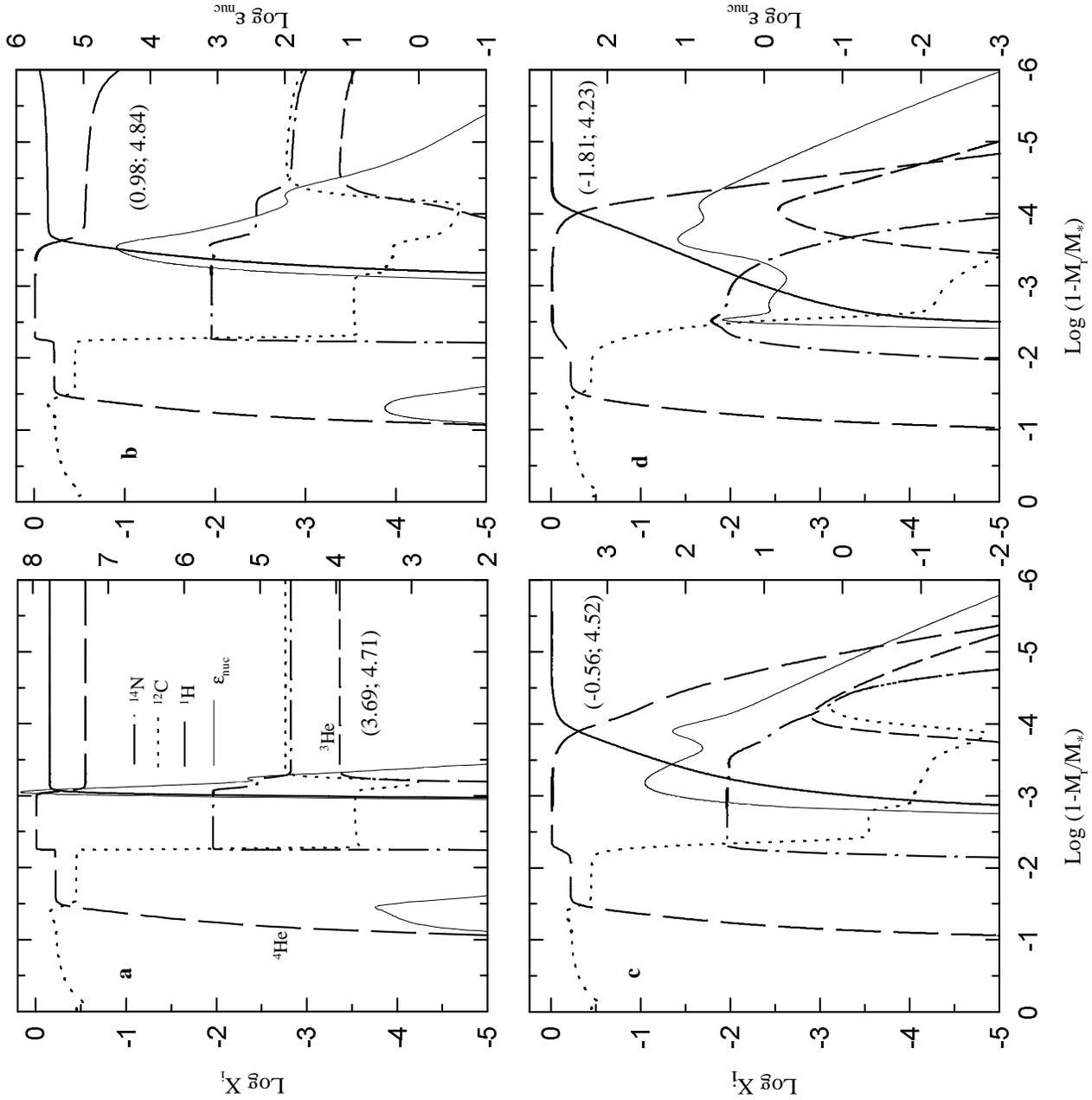} 
\caption{Abundance  by mass  of  $^1$H, $^3$He,  $^4$He, $^{12}$C  and
$^{14}$N  as  a  function  the  outer  mass  fraction  for  the  0.563
M$_{\odot}$  white  dwarf  remnant  at  selected  evolutionary  stages
characterized  by values of  $\log{L/\rm L_{\odot}}$  and $\log{T_{\rm
eff}}$  (numbers given  between brackets).   In addition,  the nuclear
energy release $\epsilon_{\rm nuc}$  (in erg/gr/s) due to hydrogen and
helium burning  is shown as thin  lines (the latter  of some relevance
only in figures  {\bf a} and {\bf b}).  The  model displayed in figure
{\bf  a} corresponds  to the  evolutionary stage  before  reaching the
point of  maximum effective  temperature at high  luminosities, whilst
the model  ploted in figure {\bf  f} corresponds to  the last computed
model.  Figure  {\bf e} depicts the  situation at the  ZZ Ceti domain.
For  this model,  note  both the  inward  extent of  the  tail in  the
hydrogen distribution caused by  chemical diffusion and the large mass
range over  which hydrogen nuclear burning  extends.  Clearly, element
diffusion  susbtantially alters  the  chemical abundance  distribution
during the white dwarf evolution.}
\end{figure*}

\setcounter{figure}{6}
\begin{figure*}
\vskip 2cm
\epsfysize=500pt
\epsfbox{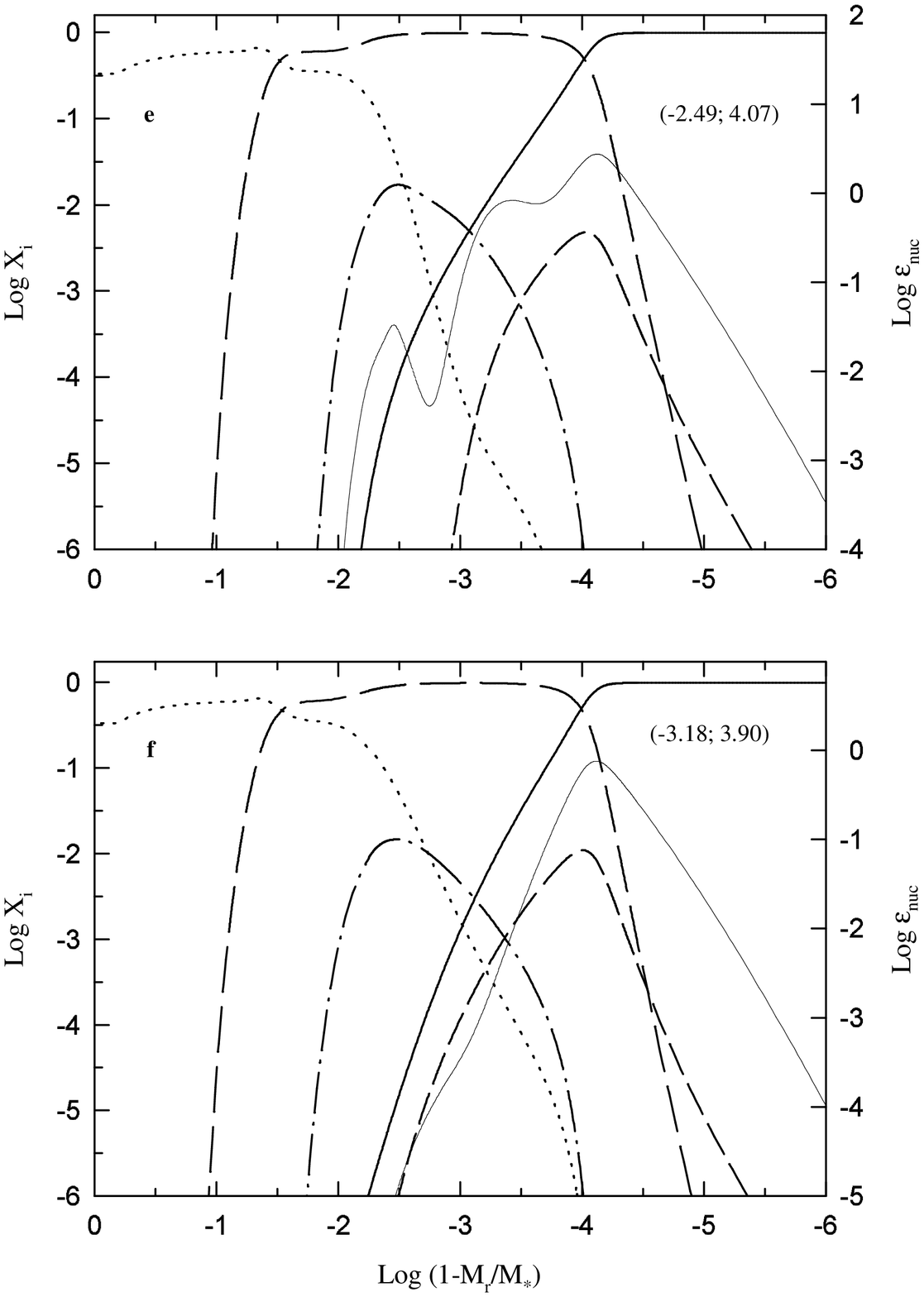} 
\caption{-Continued} 
\end{figure*}

\begin{figure*}
\vskip 5.2cm
\epsfysize=400pt
\epsfbox{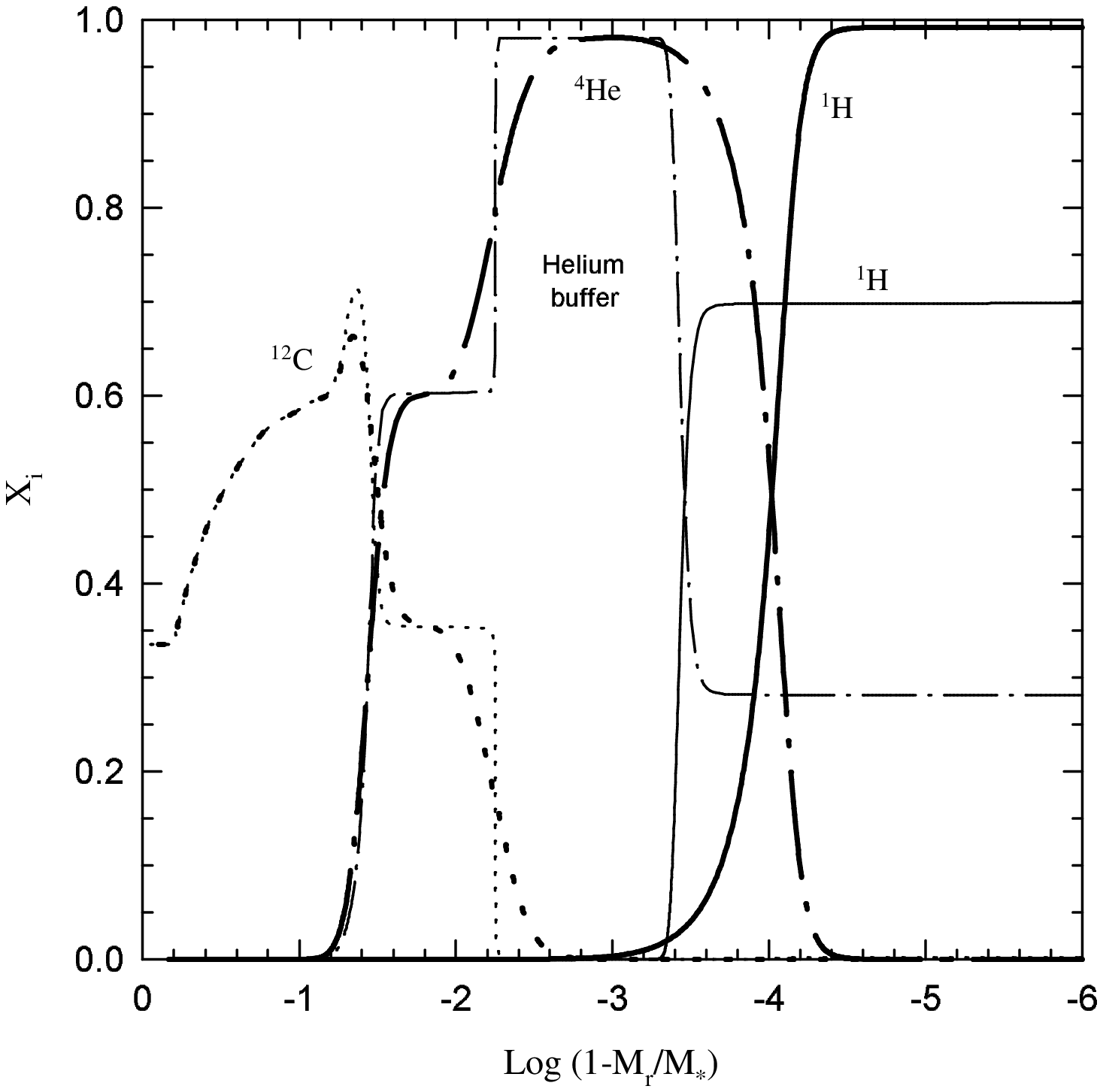} 
\caption{Abundance  profiles for our  $0.563$ M$_{\odot}$  white dwarf
remnant  for two  selected  models just  after  the maximum  effective
temperature point  (model with thin  lines) and near the  beginning of
the  ZZ  Ceti  regime.  The  models  are  characterized  by  values  (
$\log{L/\rm  L_{\odot}}$,  $\log{T_{\rm  eff}}$)  of (3.1,  5.15)  and
(-2.48, 4.07)  (thin and  thick lines, respectively).   In particular,
the distribution of hydrogen, helium and carbon (solid, dot-dashed and
dotted lines,  respectively) is  depicted as a  function of  the outer
mass fraction. The importance of element diffusion in the distribution
of chemical abundance is clearly noticeable.}
\end{figure*}

\begin{figure*}
\vskip 1cm
\epsfysize=600pt
\epsfbox{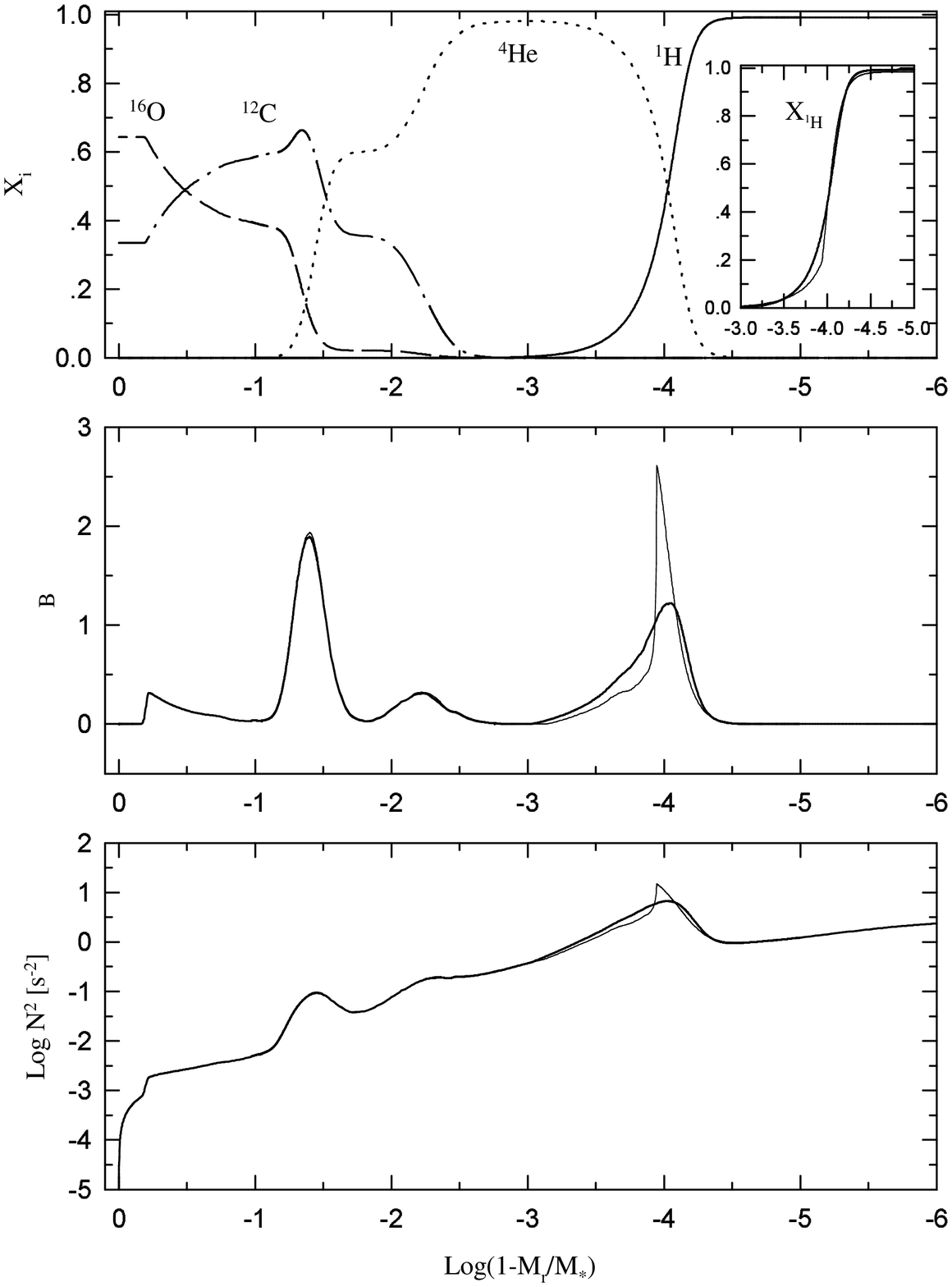} 
\caption{Upper  Panel: The internal  chemical profiles  for a  ZZ Ceti
model  of  0.563  ${\rm  M_{\odot}}$  at $T_{\rm  eff}=$  12000K,  for
hydrogen (solid line), helium  (dotted line), carbon (dot-dashed line)
and oxygen (dashed line). The  inset shows the chemical profile at the
hydrogen-helium  interface   together  with  the   prediction  of  the
diffusive equilibrium in the  trace element approximation, as given by
the thin line.  Middle panel:  the corresponding Ledoux term $B$. Thin
line depicts the results  for the diffusive equilibrium approximation.
Bottom  panel:  the   logarithm  of  the  squared  Brunt-V\"ais\"al\"a
frequency.  Thin  line  corresponds  to  the  case  of  the  diffusive
equilibrium approximation.}
\end{figure*}

\begin{figure*}
\vskip 4.2cm 
\epsfysize=480pt 
\epsfbox{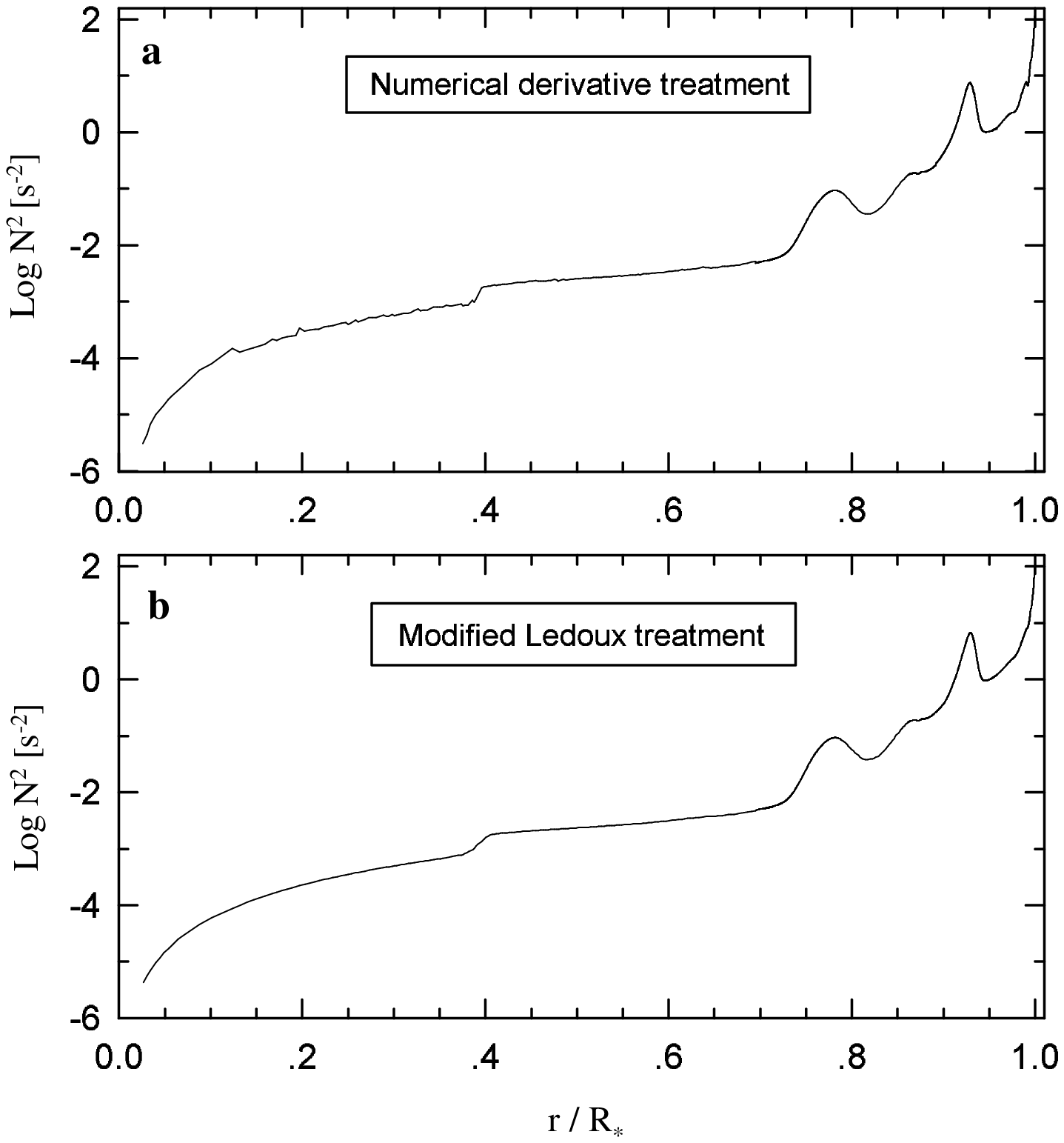}
\caption{Panel    {\bf   a}:    The   logarithm    of    the   squared
Brunt-V\"ais\"al\"a  frequency as  a  function of  the stellar  radius
computed using  numerical derivatives in Eq.   (\ref{eq5}). Panel {\bf
b}: the same  quantity but computed from Eq.  (\ref{eq2}). The stellar
model is the same as that of figure 9.}
\end{figure*}

\bsp

\label{lastpage}

\end{document}